\documentstyle[12pt,english]{article}
\newenvironment{eqn}
{\begin{equation}\begin{array}}{\end{array}\end{equation}{}}

\def\diag{{\rm diag}}
 
\def\({\Bigl(}
\def\){\Bigr)}
\def\|{\Big|}
\def\o{\circ}
\def\x{\times}

\def\ox{\otimes}

\def\pl{\oplus}

\def\SUM{\displaystyle \sum}
\def\PROD{\displaystyle \prod}
\def\BIGUPLUS{\displaystyle\biguplus}
\def\mid{\big\bracevert}

\def\then{~\Rightarrow~}
\def\sub{\subseteq}
\def\subnoteq{\subset}
\def\and{\wedge}

\def\m{\bullet}

\def\inmap{\hookrightarrow}
 
\def\A{{\,{\rm A\kern-.55emA}}}
\def\C{{\,{\rm I\kern-.55emC}}}
\def\E{{\,{\rm I\kern-.2emE}}}
\def\H{{\,{\rm I\kern-.2emH}}}
\def\I{{\,{\rm I\kern-.2emI}}}
\def\K{{\,{\rm I\kern-.2emK}}}
\def\L{{\,{\rm I\kern-.2emL}}}
\def\M{{\,{\rm I\kern-.16emM}}}
\def\N{{\,{\rm I\kern-.16emN}}}
\def\Q{{\,{\rm I\kern-.5emQ}}}
\def\R{{\,{\rm I\kern-.2emR}}}
\def\S{{\,{\rm I\kern-.42emS}}}
\def\T{{\,{\rm I\kern-.37emT}}}
\def\Z{{\,{\rm Z\kern-.35emZ}}}
\def\rin{{\,\in\kern-.42em\in}}

\def\sign{{\,{\rm sign }\,}}

\def\spec{\,{\rm spec }\,}

\def\tr{{\,{\rm tr }\,}}
\def\det{\,{\rm det }\,}

\def\centr{\,{\rm centr}\,}

\def\id{\,{\rm id}}

\def\Ad{\,{\rm Ad}\,}

\def\sup{\supseteq}
\def\supnoteq{\supset}

\def\caus{{\scriptsize{\rm caus}}}
 
\def\p{\partial}

\def\al{\alpha}  \def\be{\beta} \def\ga{\gamma}
\def\de{\delta}  \def\ep{\epsilon}  \def\ze{\zeta}
\def\th{\theta}   \def\vth{\vartheta} 
\def\ka{\kappa}   \def\la{\lambda}   \def\si{\sigma}
   \def\om{\omega} 
\def\phi{\varphi}    
    \def\La{\Lambda}

\let\rvec=\vec        % save plain's \vec as \rvec
\def\vec#1{\underline{\bf vec}_{#1}}

 \def\GL{{\bf GL}}  \def\SL{{\bf SL}}
\def\U{{\bf U}} \def\O{{\bf O}}   \def\SU{{\bf SU}} \def\SO{{\bf SO}}
  
\def \UL{{\bf UL}}  \def\D{{\bl D}}

\def\norm#1{\parallel #1\parallel}
\def\d#1{\check{#1}}
\def\angle#1{\langle#1\rangle}

\def\brack#1{\lbrack#1\rbrack}

\def\Brack#1{\Bigl\lbrack#1\Bigr\rbrack}

\def\ol#1{\overline{#1}}
\def\bl#1{{\bf #1}}
\def\cl#1{{\cal #1}}

\def\ro#1{{\rm #1}}

\def\dprod#1#2{\langle#1,#2\rangle}
\def\sprod#1#2{\langle#1|#2\rangle}
\def\com#1#2{\lbrack#1,#2\rbrack}

\def\acom#1#2{\{#1,#2\}}

\def\map{\longrightarrow}

\def\mape{\longmapsto}

\def\dmap{\Big\downarrow}

\def\Diagr#1#2#3#4#5#6#7#8{\matrix{\noalign{\vskip5mm}
      &              &{\scriptstyle #5}&              &     \cr
      & #1           & \map           & #2           &     \cr
{\scriptstyle #8}   &\dmap         &    &\dmap  &{\scriptstyle#6} \cr
      & #4           & \map           & #3           &     \cr
      &              &{\scriptstyle#7}&              &     \cr
\noalign{\vskip5mm}             }}

\def\Time{\ro{{\scriptsize time}}}
\def\Light{\ro{{\scriptsize light}}}
\def\Space{\ro{{\scriptsize space}}} 
\def\caus{\ro{{\scriptsize causal}}} 
\def\meas{\ro{meas}}

\begin{document}

\begin{titlepage}

\hfill MPI-PhT/97-42
\vskip3cm
\centerline{\bf REALIZATIONS OF CAUSAL MANIFOLDS}
\vskip5mm
\centerline{\bf BY QUANTUM FIELDS}
\vskip2cm
\centerline{
Heinrich Saller\footnote{\scriptsize 
e-mail adress: saller@mppmu.mpg.de} 
}
\centerline{Max-Planck-Institut f\"ur Physik and Astrophysik}
\centerline{Werner-Heisenberg-Institut f\"ur Physik}
\centerline{M\"unchen}
\vskip25mm

\centerline{\bf Abstract}
Quantum mechanical operators and quantum fields are 
interpreted as realizations
of timespace manifolds. 
Such causal manifolds are parametrized by 
the classes of the positive unitary operations in all complex operations, i.e.
by the homogenous
spaces $\D(n)=\GL(\C^n_\R)/\U(n)$ with $n=1$ for mechanics and $n=2$ for
relativistic fields. The rank $n$ gives the number of both the discrete and
continuous invariants used in the harmonic analysis,
i.e. two characteristic masses in the relativistic case.
 'Canonical' field theories
with the familiar divergencies are inappropriate realizations of the 
real 4-dimensional causal
manifold $\D(2)$. Faithful timespace realizations 
do not lead to divergencies. In general they are reducible, but
nondecomposable - in addition to representations with eigenvectors
(states, particle) they incorporate principal
vectors without a particle (eigenvector) basis as exemplified by the
Coulomb field.

\end{titlepage}

\newpage

\tableofcontents

\newpage

\advance\topmargin by -1.6cm

\centerline{\bf Introduction}

Quantum theory deals primarily with operations - e.g. 
ti\-me\-spa\-ce translations and rotations or hypercharge and 
isospin transformations. Its experimental interpretation relies on states
or particles, i.e. eigenvectors of asymptotically relevant operations.
The particle properties 
(mass, spin, charge etc.) we are measuring are the corresponding eigenvalues.
The modality structure of quantum theory, e.g. the probability amplitudes,
is formulated using complex linear structures with conjugations. 
Therefore, complex
linear operations and their unitary substructures - not necessarily positive
unitary\footnote{\scriptsize
In the {\it orthogonal} and {\it unitary} groups
$\O(N_+,N_-)$ and $\U(N_+,N_-)$ resp. the {\it positive} orthogonal and unitary ones 
are $\O(N)$ and $\U(N)$ resp.} 
 - play the central role in quantum theory.

The characterization of particles as positive unitary irreducible
re\-pre\-sen\-ta\-tion of the Poincar\'e group by Wigner \cite{WIG} is an asymptotic
experimentally oriented language.
A particle language, i.e. an eigenvector basis,
 is too narrow to describe  interactions.
Following up the identity of particles in their interaction is impossible - 
one tries
to evade or circumvent the particle-interaction complementarity 
using words and concepts like 'off shell' or 
'virtual particles' or 'ghosts' etc. There are  interaction structures    
which do not show  up in the projection to a particle basis. 

First of all, the nontranslative ho\-mo\-ge\-neous interaction symmetries
are truly larger than particle symmetries. Concerning
external operations, particles have ho\-mo\-ge\-neous symmetry properties only with
respect to the 'little groups' spin $\SU(2)$ or helicity (polarization)
$\U(1)$, which are true subgroups of the interaction compatible 
Lorentz $\SL(\C^2_\R)$ group. With respect to 
the internal operations as seen  in the interactions of the 'standard model of 
elementary particles',  there  remains for particles only an abelian  
$\U(1)$ electromagnetic symmetry, defined as the internal 
'little group' (fixgroup) of the 
degenerated  ground state, as subgroup of the hyperisospin transformation group
$\U(2)$.
If colour $\SU(3)$ confinement holds, only colour
singlets arise as particles. 

In addition to these projections from the 'large' interaction symmetries
to the 'little' particle symmetries,
both for external and internal operations, there are operational structures 
in the dynamics without
any asymptotic particle trace, the most prominent ones given by the Coulomb
and gauge degrees of freedom of the standard interactions, formalized in 
relativistic nonabelian quantum field theory in cooperation with 
Fadeev Popov degrees of
freedom \cite{BRS,NAK}. All those 'ghost' structures come in connection with 
{\it indefinite unitary}
 re\-pre\-sen\-ta\-tions of ti\-me\-spa\-ce translations \cite{S96}, 
ultimately tied to the indefinite structure of the
noncompact Lorentz group $\O(1,3)$ which is indispensible for a 
nontrivial relativistic causal 
order.

This paper is an attempt to  connect the asymptotic
concepts 'particle' and even 'time' and 'space' 
with the interactions on a deeper operation oriented \cite{FI}
level. Timespace will be formalized by a coset structure
- as the noncompact real ho\-mo\-ge\-neous space 
$\D(n)=\GL(\C^n_\R)/\U(n)$
which is the manifold of the  positive unitary, probability amplitude related 
operations $\U(n)$ in  all complex linear ones. The causal 
manifold $\D(n)$ has real dimension $n^2$ and real rank $n$. The abelian case $n=1$ 
involves the real 1-di\-men\-sio\-nal
causal group (time) $\D(1)$ as the  framework which is extensively used in
quantum mechanics. The next simple case with ti\-me\-spa\-ce rank $n=2$, 
the first nonabelian one, involves 
the real 4-di\-men\-sio\-nal ho\-mo\-ge\-neous manifold $\D(2)$ with internal stability
group $\U(2)$, it shows all the features familiar 
from relativistic quantum field theories.

The rank $n$ of the causal manifold $\D(n)$ shows up in the number of 
real continuous invariants used
in its re\-pre\-sen\-ta\-tions. The re\-pre\-sen\-ta\-tions 
of the causal group $\D(1)$ are characterized by one 
continuous invariant (frequency)
$\om$,
which serves e.g. as unit in energy eigenstates of a quantum mechanical
dynamics,  
whereas the re\-pre\-sen\-ta\-tions of $\GL(\C^2_\R)$ for
the ho\-mo\-ge\-neous ti\-me\-spa\-ce $\D(2)$ 
involve two 
continuous invariants 
which may be tentatively called a particle mass $m_1$ 
and an interaction mass
$m_2$ - or $m_1$ and a dimensionless ratio ${m_2^2\over m_1^2}$, possibly
related to the coupling constants used  for relativistic interactions.

A parametrization of ti\-me\-spa\-ce manifold $\D(n)$ realizations
by vectors of  complex linear re\-pre\-sen\-ta\-tion spaces
leads to the concepts of quantum mechanical operators 
for $n=1$ and relativistic 
quantum fields for $n=2$. In the latter case, the re\-pre\-sen\-ta\-tion of the
rank 2
operations in $\D(2)$ - not only rank 1 (time translations)
 as done in conventional particle oriented linear quantum
fields - gives rise to a framework where product re\-pre\-sen\-ta\-tions
are definable without the light cone supported divergencies found for 
interacting linear quantum fields.

%\newpage

\section{'Canonical' Quantization}

Quantum mechanics as
a theory for time dependent operators
was very  successfull. The quantization involved,  called 'canonical', 
was 
simply taken over - in a Lorentz compatible extension - 
for ti\-me\-spa\-ce dependent operators (quantum fields).

\subsection{Quantum Structure in Mechanics}

Heisenberg's noncommutativity condition $\com{iP}X=1$ 
(in units with $\hbar=1$) for 
position-momentum operator pairs $(X,P)$ is 
the trivial time $ t=0$ element of  time  
re\-pre\-sen\-ta\-tions\footnote{\scriptsize
The linear dependence is used in the notation
$\com{a(y)}{b(x)}=\com ab(x-y)$ etc.} $\com{iP}X( t)$
 by a
quantum mechanical dynamics, e.g. in the 
- not only historically relevant - simplest example of
the harmonic oscillator with Hamiltonian 
$H={P^2\over 2M}+\ka{X^2\over2}$, involving a
mass $M$ and a spring constant $\ka$ or a frequency  $m^2={\ka\over M}$
and an intrinsic length $\ell^4={1\over\ka M}$ 
\begin{eqn}{l}
\left.\begin{array}{l}
{d\over d t}X( t)=m\ell^2P( t)\cr 
{d\over d t}P( t)=-{m\over\ell^2} X( t)\end{array}\right\} 
\then \com{iP}{X}( t)=\cos m t
\end{eqn}The eigenvectors of the time translations can be built by
products of the cre\-ation-an\-ni\-hi\-la\-tion pair $(\ro u,\ro u^\star)$ 
\begin{eqn}{l}
\left.\begin{array}{rl}
\ro u&={1\over\ell\sqrt2}X-{\ell\over\sqrt2}iP\cr
\ro u^\star&={1\over\ell\sqrt2}X+{\ell\over\sqrt2}iP\cr
\end{array}\right\}\then\left.\begin{array}{l}
{d\over d t}(\ro u^\star)^k(\ro u)^l( t)
=i(l-k)m(\ro u^\star)^k(\ro u)^l( t)\cr
\hbox{with }k,l\in\N\end{array}\right.
 \end{eqn}
 
 The harmonic oscillator as $\U(1)$-time re\-pre\-sen\-ta\-tion 
 uses a point measure for the
frequencies as the linear forms on the  time translations
\begin{eqn}{rl}
\U(1)\ni\com{\ro u^\star}{\ro u}( t)&=e^{im t}\cr
&=\int d\mu  \de(\mu -m)e^{i \mu t }\cr
&={\ep( t)\over i\pi}\int d\mu {\mu + m\over \mu^2- m^2_P}  e^{i \mu t }=
{1\over 2i\pi }\oint {d\mu\over \mu- m}e^{i \mu t } \cr
\end{eqn}Here $m_P$ denotes the principal 
value integration and $\oint$
the mathematically positive circle around all poles in the complex 
frequency plane $\mu\in\R\subnoteq\C$.

The selfdual time re\-pre\-sen\-ta\-tion\footnote{\scriptsize
The $\SO(2)$-matrices 
${\scriptsize\pmatrix{
\cos m t&e^\al\sin m t\cr 
-e^{-\al}\sin m t&\cos m  t\cr}}$ with $\al\in\C$ are equivalent to the
familiar $\SO(2)$-matrices with $\al=0$.}
in $\SO(2)$ by position and momentum 
reads 
\begin{eqn}{rl}
\SO(2)&\ni
{\scriptsize\pmatrix{
\com{iP}X& \com{X}X\cr
\com{P}P&\com{X}{-iP}\cr}}( t)
={\scriptsize\pmatrix{
\cos m t&i\ell^2\sin m t\cr 
{i\over\ell^2} \sin m t&\cos m  t\cr}}\cr
{\scriptsize\pmatrix{
\cos m t\cr i\sin m t\cr}} 
&=\int d\mu \de(\mu^2- m^2)
\ep(\mu){\scriptsize\pmatrix{
\mu\cr m\cr}}e^{i\mu t }
=\int d\mu \de(\mu^2- m^2)
\ep( m){\scriptsize\pmatrix{
 m\cr\mu\cr}}e^{i \mu t }\cr
&={\ep( t)\over i\pi}\int {d\mu\over
\mu^2- m_P^2}
{\scriptsize\pmatrix{\mu\cr m\cr}} e^{i \mu t }
={1\over i\pi}\oint {d\mu \over \mu^2- m^2}
{\scriptsize\pmatrix{\mu\cr m\cr}}e^{i \mu t }\cr
\end{eqn}

In 
general, a dynamics is not linear,
i.e. the Hamiltonian is not quadratic in the fundamental operator pairs $(X,P)$,
 e.g. for the hydrogen atom with
the  Hamiltonian $H={\rvec P^2\over 2M}-{g_0\over |\rvec X|}$.
In those 'truly interacting' cases, the operators for the
energy eigenstates may be
complicated combinations of positions $X$ and 
momenta $P$.

\subsection{Distributive Quantization of Particle Fields}

The relativistic embedding of the $\SO(2)$ time 
re\-pre\-sen\-ta\-tions leads to two different results: Since the energy is
embedded in a Lorentz energy-momentum vector
$\mu \inmap (q^k)_{k=0}^3$,
one obtains both 'scalar and vector 
cosinus and sinus'
\begin{eqn}{l}
{d\over d t}{\scriptsize\pmatrix{
 \cos m t\cr i\sin m t \cr}}=im{\scriptsize\pmatrix{
 i\sin m t \cr \cos m t\cr}}\inmap\cases{
\p_k
{\scriptsize\pmatrix{
 \bl c^k(m|x)\cr i\bl s(m|x)\cr}}
=im{\scriptsize\pmatrix{
 i\bl s(m|x)\cr \bl c_k(m|x)\cr}}\cr
\p_k
{\scriptsize\pmatrix{
 \bl C(m|x)\cr i\bl S^k(m|x)\cr}}
=im{\scriptsize\pmatrix{
 i\bl S_k(m|x)\cr \bl C(m|x)\cr}}\cr}
\end{eqn}which both fulfill a
ho\-mo\-ge\-neous  Klein-Gordon equation
\begin{eqn}{l}
({d^2\over d t^2}+m^2)
{\scriptsize\pmatrix{
 \cos m t\cr i\sin m t\cr}}=0
 \inmap\cases{
 (\p^2+m^2)
{\scriptsize\pmatrix{
 \bl c^k(m|x)\cr i\bl s(m|x)\cr}}=0\cr
 (\p^2+m^2)
{\scriptsize\pmatrix{
 \bl C(m|x)\cr  i\bl S^k(m|x)\cr}}=0\cr}\cr
   \end{eqn}

The embedding with an ordered
Dirac energy-momentum measure at  $q^2=m^2$
for the translation eigenvectors $e^{iqx}$
\begin{eqn}{l}
{\scriptsize\pmatrix{
\bl c^k(m|x)\cr
i\bl s(m|x)\cr}}
=\int {d^4 q\over(2\pi)^3}\de(q^2-m^2)\ep(q_0)
{\scriptsize\pmatrix{
q^k\cr m\cr}} e^{ixq}=
{\ep(x_0)\over i\pi }\int {d^4 q\over (2\pi)^3}{1\over q^2-m^2_P}
{\scriptsize \pmatrix{ q^k\cr m\cr}}e^{ixq}\cr
\end{eqn}defines the 
causally supported quantization distributions
\begin{eqn}{l}
{\scriptsize\pmatrix{
\bl c^k(m|x)\cr
i\bl s(m|x)\cr}}=0\hbox{ for spacelike } x^2<0\cr
\end{eqn}

The embedding with a 'not ordered' 
measure
\begin{eqn}{l}
{\scriptsize\pmatrix{
\bl C(m|x)\cr
i\bl S^k(m|x)\cr}}
=\int {d^4 q\over(2\pi)^3}\de(q^2-m^2)\ep(m)
{\scriptsize\pmatrix{
m\cr q^k\cr}} e^{ixq}\cr
\end{eqn}defines the Fock state functions which have also nontrivial
spacelike contributions.

To illustrate  quantum fields
with linear equations of motion and particle interpretation,
following the harmonic oscillator linear  
quantum structures,  a free Dirac field $\bl\Psi$
for particles with mass $m\ne0$, e.g. the electron,
yields a good example
\begin{eqn}{l}(i{d\over d t}+m)\ro u( t)=0\inmap
(i\ga_k\p^k+m)\bl\Psi(x)=0,~~
\end{eqn}Weyl fields arise for $m=0$ with a left or right handed 
Weyl re\-pre\-sen\-ta\-tion 
${1+\ga_5\over2}\ga_k\cong\rho_k=(\bl 1,\rvec\si)$
and ${1-\ga_5\over2}\ga_k\cong\d\rho_k=(\bl 1,-\rvec\si)$ resp.
replacing the Dirac re\-pre\-sen\-ta\-tion. 

The Feynman propagator is the Fock value $\angle{...}$
of the sum of the causally ordered quantization
an\-ti\-com\-mu\-ta\-tor  and the commutator
\begin{eqn}{rl}
\angle{\cl C\ol{\bl\Psi}\bl\Psi }(x)
&={i\over\pi}\int {d^4q \over(2\pi)^3}{\ga_kq^k+m\over
q^2-m^2+io}e^{iqx}\cr
&=\angle{-\ep(x_0)\acom{\ol{\bl\Psi}}{\bl\Psi}(x) 
+ \com{\ol{\bl\Psi}}{\bl\Psi }(x)}\cr
\end{eqn}The Fock form function with its spacelike contributions 
\begin{eqn}{rl}
\angle{\com{\ol{\bl\Psi}}{\bl\Psi }}(x)
&=\bl {Exp}(im|x)=
\bl C(m|x)+i\ga_k\bl S^k(m|x)\cr
&=\int {d^4 q\over (2\pi)^3}\ep(m)(\ga_kq^k+m)\de(q^2-m^2)e^{ixq}\cr
\end{eqn}will be discussed later (subsection 5.3).
The distribution for the quantization anticommutator
\begin{eqn}{rl}
\acom{\ol{\bl\Psi}}{\bl\Psi}(x)
&=\bl {exp}(im|x)=
\ga_k\bl c^k(m|x)+i\bl s(m|x)\cr
&=\int {d^4 q\over (2\pi)^3}\ep(q_0)(\ga_kq^k+m)\de(q^2-m^2)e^{ixq}\cr
&={\ep(x_0)\over i\pi }\int {d^4 q\over (2\pi)^3}{\ga_kq^k+m\over q^2-m^2_P}
e^{ixq}\cr
\end{eqn}is the
interaction relevant structure,
which causes the 
'divergencies' in relativistic field theories (not
the particle relevant Fock value of the commutator).
The distinction between
$\bl{exp}(im|x)$ and $\bl{EXP}(im|x)$ is characteristic for
the relativistic case with its causal partial order.

The quantization distribution is  given explicitely in ti\-me\-spa\-ce coordinates 
\begin{eqn}{rccl}
\bl c^k(m|x)&={m^2\over8\pi}&
{m^2\ep(x_0)x^k\over2}&
\brack{\de'({m^2x^2\over4})-\de({m^2x^2\over4})+{1\over2}\vth(x^2)
\cl E_2(m^2x^2)}\cr
\bl s(m|x)&={m^2\over8\pi}&
m\ep(x_0)&\brack{\de({m^2x^2\over4})
-\vth(x^2) \cl E_1(m^2x^2)} \cr
\end{eqn}with $\bl c^k(0|x)={\ep(x_0)x^k\over\pi}\de'(x^2)$.
These distributions involve the 
functions $\cl E_n$, in general for $n=0,1,\dots$ with $\tau^2=
\vth(x^2) m^2x^2$
('eigentime' $\tau$),
Bessel coefficients $J_n$  \cite{SNED} and the beta-function $B$
\begin{eqn}{l}
 \cl E_n(\tau^2)=n!{J_n(\tau)\over({\tau\over2})^n}
=n!{\SUM_{j=0}^\infty}{ (-{\tau^2\over4})^j\over j!(j+n)!}
={\int_0^1 d\mu^2 (\mu^2)^{-{1\over2}} (1-\mu^2)^{-{1\over2}+n}\cos\mu\tau
\over B({1\over2},{1\over2}+n)}\cr
\cl E_{n+1 }(\tau^2)=-(n+1 ){d \cl E_n(\tau^2)\over d{\tau^2\over4}}
,~~\cl E_n(0)=1\cr
\end{eqn}

The vectorial distribution for time $x_0=0$ describes the 
quantization of linear fields
\begin{eqn}{l}
\bl{exp}(im| x)|_{x_0=0}=\ga_k\bl c^k(m|\rvec x)=\ga_0\de^3(\rvec x ),~~
\int d^3 x\acom{\ol{\bl\Psi}}{\bl\Psi}(x)|_{x_0=0}=\ga_0
\end{eqn}The fact, that the linear quantization gives 
no functions, e.g.
that $\bl{exp}(im|0)$  does not make sense 
 or that two quantizations cannot be simply multiplied   
 for a product re\-pre\-sen\-ta\-tion
of the translations, e.g. $\bl{exp}(im|x)\bl{exp}(im|x)$ 
arising in a perturbative approach
(vacuum polarization in quantum electrodynamics), 
illustrates the familiar divergence problem for
linear fields which has 
to be treated by sophisticated techniques. It hints to the
inappropriateness of particle fields for  
more than a perturbative formulation
of interactions,  e.g. their inappropriateness 
to parametrize  a bound state problem. 

Using a decomposition of all translations into time and space translations,
e.g. given in a massive particle rest system,
the time ordered integral 
of the 
spacelike trivial quantization distributions gives
- after interchanging time and energy integration -  
the space dependent interaction functions, e.g. the Yukawa interaction 
and force 
\begin{eqn}{rll}
{1\over2}\int dx_0 \ep(x_0)\bl s(m|x)&=md(m,|\rvec x|)
=m\int {d^3q\over(2\pi)^3}
{e^{-i\rvec q\rvec x}\over \rvec q^2+m^2}
&={m\over 4\pi|\rvec x|}e^{-m|\rvec x|}\cr
{1\over2}\int dx_0 \ep(x_0)\ga_k\bl c^k(m|x)&=\ga_a\p^a d(m,|\rvec x|)
&=-\rvec\ga\rvec x ~{1+m|\rvec x|\over 4\pi|\rvec x|^3} e^{-m|\rvec x|}\cr
\end{eqn}The singularities at $\rvec x=0$
reflect the divergency problems of linear fields, if they are used to
parametrize interactions. 

Interchanging space and momenta integrations, the 
space integrals of the 
quantization distributions 
lead to the time development of the harmonic oscillator 
\begin{eqn}{l}
\int d^3x~\ga_k\bl c^k(m|x)=\ga_0\cos mx_0,~~
\int d^3x~\bl s(m|x)=\sin mx_0
\end{eqn}

The distributive quantization of relativistic fields
identifies the eigenvalue of the harmonic oscillator 
time translation dependence, given by the frequency $m_0$, with 
the characteristic mass for the  Yukawa interaction space   
dependence, given by the inverse range $m_r$
\begin{eqn}{l}
({d^2\over dt^2}+m_0^2)\cos m_0t=0,~~ 
(-{\p^2\over \p\rvec x ^2}+m_r^2)d(m_r,|\rvec x|)=\de(\rvec x)
\end{eqn}in the Lorentz compatible ti\-me\-spa\-ce translation dependence
\begin{eqn}{l} 
\hbox{with }m_0=m_r:~~(\p^2+m^2)\bl c^k(m|x)=0
\end{eqn}

\section{Causal Timespaces}

Observable in mechanics depend on  time coordinates,
 relativistic fields depend on ti\-me\-spa\-ce coordinates. In this 
section, 
ti\-me\-spa\-ces are
reformulated in a general   symmetry oriented algebraic framework
starting from complex linear operations.

\subsection{Causal Complex Algebras}

Cartan's re\-pre\-sen\-ta\-tion of the ti\-me\-spa\-ce translations 
(real 4-di\-men\-sio\-nal Min\-kow\-ski vector space) uses
hermitian complex $2\x 2$-matrices from the 
associative unital algebra 
of all complex $2\x2$-matrices. Together with the  
time translations (real numbers) of mechanics, they 
are given as follows   
\begin{eqn}{l}
x=\left\{\begin{array}{cl}
t=\ol t&\in\C(1)\cr
{1\over2}{\scriptsize\pmatrix{
x_0+x_3&x_1-ix_2\cr x_1+ix_2&
x_0-x_3\cr}}=x^\star&\in\C(2)\end{array}\right.
\end{eqn}Those familiar cases should be used as illustrations for 
the general case working with  complex $n\x n$  matrices for $n\ge1$.
 
The matrices $z\in \C(n)$ with the
usual hermitian conjugation $\star$ are an involutive algebra, decomposable
into two isomorphic real vector spaces of dimension  $n^2$
\begin{eqn}{rl}
\C(n)&= \R(n)\pl i\R(n)\cong\R^{2n^2},~~z=x+iy\cr
\R(n)&=\{x\in\C(n)\mid x=x^\star\}\cong\R^{n^2}\cr
\end{eqn}The symmetric vector subspace  will be called the
{\it matrix re\-pre\-sen\-ta\-tion $\R(n)$ of the ti\-me\-spa\-ce translations}
with $n$ the {\it rank of ti\-me\-spa\-ce}.

A  basis for $\C(n)$ is given by
\begin{eqn}{l}
\{\rho(n)^j,i\rho(n)^j\}_{j=0}^{n^2-1}\hbox{ with }
\{\rho(n)^j\}_{j=0}^{n^2-1}
=\{{\bl 1(n)\over n},{\si(n)^a\over2}\}_{a=1}^{n^2-1}\cr
z=z_j\rho(n)^j,~~z_A^{\dot A}=z_j\brack{\rho(n)^j}_A^{\dot A},~~
A,\dot A=1,\dots,n\cr
\end{eqn}using the unit matrix $\bl 1(n)$ and $(n^2-1)$ generalized hermitian
traceless Pauli matrices $\si(n)^a$ for $n\ge2$, i.e.   
three Pauli matrices $\rvec \si$  for $n=2$, eight Gell-Mann matrices 
$\si(3)^a=\la^a$ for $n=3$ etc. 
\begin{eqn}{ll}
\si(n)^a=\si(n)^{a\star},~~\tr \si(n)^a=0&\cr
\com{i\si(n)^a}{i\si(n)^b}=\al^{abc}i\si(n)^c,&
\hbox{totally antisymmetric }\al^{abc}\in\R\cr
\acom{\si(n)^a}{\si(n)^b}=2\de^{ab}\bl 1(n)+\de^{abc}\si(n)^c,&
\hbox{totally symmetric }\de^{abc}\in\R\cr
\end{eqn}

The determinant defines the {\it abelian projection}
of $\C(n)$ to the complex numbers $\C(1)=\R\pl i\R$, compatible
with the unital multiplication and the conjugation
(a $\star$-monoid morphism) 
\begin{eqn}{l}
\det:\C(n)\map\C,~~z\mape z^n=\det z,~~\cases{
\det z\o z'=\det z~\det z'\cr
\det \bl 1(n)= 1\cr
\det z^\star=\ol{\det z}\cr}
\end{eqn}By polarization, i.e. by 
combining appropriately  $(z_1\pm z_2\pm\dots\pm z_n)^n$,
one obtains a totally symmetric multilinear form, familiar 
for the Min\-kow\-ski translations $\R(2)$ as Lorentz bilinear form
with indefinite signature 
\begin{eqn}{rl}
\eta:&\C(n)\x\cdots\x\C(n)\map\C\cr
&(z_1,\dots,z_n)\mape
\eta(z_1,\dots,z_n)=\ep^{A_1\dots A_n}\ep_{\dot A_1\dots \dot A_n}
(z_1)_{A_1}^{\dot A_1}\cdots(z_n)_{A_n}^{\dot A_n} \cr
n=1:&~\eta (z)=\det z= z\cr
n=2:&~\eta(z_1,z_2)=(z_1+z_2)^2-(z_1-z_2)^2,~~
\sign\eta|_{\R(2)}=(1,3)\cr
\end{eqn}

The trace part and the traceless part of a translation will be 
called a time translation in $\T\cong\R$ and a space
translation in $\S(n)\cong\R^{n^2-1}$
\begin{eqn}{l}
x=x_j\rho(n)^j=x_0{\bl 1(n)\over n}+ x_a {\si(n)^a\over2}
\end{eqn}However, a  decomposition into time and space translations is 
incompatible with the determinant since 
for $n\ge2$ in general $\det(x+y)\ne \det x+\det y$.

The involutive algebra $\C(n)$ is a C$^*$-algebra
 \cite{BRAROB,RIC}with the norm $\norm z^2=\sprod zz$ 
induced by the scalar product
\begin{eqn}{l}
z,z'\in\C(n)\then \sprod z{z'}=\tr z'\o z^\star
\end{eqn}Any C$^*$-algebra is (partially) ordered via the spectrum 
\begin{eqn}{l}
x\succeq 0\iff x=x^\star\hbox{ and }\spec x\subnoteq
 \R^+=\{\al\in\R\mid\al\ge0\}\cr
\end{eqn}Therewith, $\C(n)$ will be called a {\it causal complex algebra}.

All ti\-me\-spa\-ce translations can be 
diagonalized
 with a real spectrum
\begin{eqn}{rl}
x\in \R(n)&\then\spec x=\{\xi\mid\det(x-\xi\bl 1(n))=0\}\subnoteq \R\cr
n=1:~~&x\in\R(1),~~\spec x=\{\tau\}\cr
n=2:~~&x\in\R(2),~~\spec x=\{{x_0\pm|\rvec x|\over2}\}\cr
\end{eqn}The $n$ real spectral values $\{\xi_r\}_{r=1}^n$ of a ti\-me\-spa\-ce
translation $x\in\R(n)$  will be called  its {\it Cartan 
coordinates}.

A linear transformation $z\in\C(n)$ is diagonalizable  
$z=u\o\diag(z)\o u^{-1}$ if, 
and only if, it is normal $z\o z^\star= z^\star\o z$. 
The diagonalization transformation is
unitary $u^{-1}=u^\star$. Therewith $x\in\R(n)$ and $l\in i\R(n)$ are
diagonalizable, but not any $z\in\C(n)$.
 
The C$^*$-algebra order generalizes the familiar order of the 1- 
and 4-di\-men\-sio\-nal
translations. 
With one nontrivial positive 
causal vector $c$, positivity is  expressable
by positive $c$-projected products ($n$ causal projections)
\begin{eqn}{l}
c,x\in\R(n),~~c\succ0\cr
x\succeq 0\iff x^r_c=
\eta(\underbrace{x,\dots,x}_r,\underbrace{c,\dots,c}_{n-r})\ge0
\hbox{ for }r=1,\dots,n
\end{eqn}

The characteristic functions for the  causal translations
use the spectral values 
\begin{eqn}{rl}
x\in\R(n),~~
x^n=\det x&={\PROD_{r=1}^n}\xi_r,~~\xi_r\in\spec x\cr
\vth(x)&={\PROD_{r=1}^n}\vth(\xi_r)={\PROD_{r=1}^n}\vth(x^r_c),~~c\succ 0\cr
\ep(x)&=\vth(x)-\vth(-x)\cr
\end{eqn}with the familiar  example
for  Min\-kow\-ski ti\-me\-spa\-ce with a time component
$\eta(x,c)=x_0$ 
\begin{eqn}{l}
n=2:~~\cases{
\vth(x)=\vth(x_0+|\rvec x|)\vth(x_0-|\rvec x|)=\vth(x_0)\vth(x^2)\cr
\ep(x)=\ep(x_0)\vth(x^2)\cr}
\end{eqn}

The vector space of all ti\-me\-spa\-ce translations $\R(n)$
is the union of 
the  positive and the negative {\it causal cone} and the
{\it spacelike} submanifold
\begin{eqn}{rl}
\R(n)&=\R(n)_\caus\cup \R(n)_\Space,~~
\R(n)_\caus\cap \R(n)_\Space=\{0\}\cr
\R(n)_\caus&=\R(n)_\caus^+ \cup\R(n)_\caus^-,~~
\R(n)_\caus^+ \cap\R(n)_\caus^-=\{0\}\cr
\R(n)_\caus^+ &=\{x\in\R(n)\mid \spec x\subnoteq\R^+\}=-\R(n)_\caus^- \cr
\end{eqn}All translations can be written as sum of a positive and a negative
causal translation \begin{eqn}{l}
\R(n)=\{x_+ +x_-\mid x_+,-x_-\in \R(n)_\caus^+ \}
\end{eqn}

The positive causal  cone is the disjoint union 
\begin{eqn}{l}
\R(n)_\caus^+ =\{0\}\uplus\R(n)^+_\Time\uplus \R(n)^+_\Light
\end{eqn}of the trivial
translation (tip of the cone), 
the strictly positive {\it timelike} translations (interior of the cone)
where the spectrum does not contain  0 
\begin{eqn}{l}
\R(n)^+_\Time=\{x\in\R(n)_\caus^+ \mid 0\not\in\spec x\}
\end{eqn}and the strictly positive {\it lightlike} translations (skin of the 
tipless cone) where  0
is a spectral value
\begin{eqn}{l}
\R(n)^+_\Light=\{x\in\R(n)_\caus^+ \mid x\ne 0,~~ 0\in\spec x\}
\end{eqn}

For the 1-di\-men\-sio\-nal totally ordered translations $\R(1)=\R$,
one has a trivial space
$\R(1)_\Space=\{0\}$.
The nontrivial spacelike manifold for $n\ge2$ is the disjoint union of
$(n-1)$ submanifolds $\R(m,n-m)_\Space$ with $m$ strictly positive 
and $n-m$ strictly negative Cartan coordinates
\begin{eqn}{l}
n\ge2:~~\R(n)_\Space\setminus\{0\}={\BIGUPLUS_{m=1}^{n-1}}\R(m,n-m)_\Space
\end{eqn}

In the 1-di\-men\-sio\-nal case 
there is no  light $\R(1)_\Light^\pm=\emptyset$. 
Light is a genuine nonabelian
phenomenon, arising for $n\ge2$.
There, the strictly
positive (negative) lightlike manifold is the  
disjoint union of $(n-1)$
submanifolds $\R(m,n-1-m)^\pm _\Light$ with exactly $m$  trivial 
and $n-1-m$ strictly positive (negative) Cartan coordinates  
\begin{eqn}{l}
n\ge2:~~\R(n)^\pm_\Light={\BIGUPLUS_{m=1}^{n-1}}\R(m,n-1-m)^\pm_\Light
\end{eqn}

The {\it proper time ('eigentime') projection} is the  causal projection
of the translations to the real numbers $\R(1)=\R$
\begin{eqn}{l}
\tau:\R(n)\map\R,~~\tau(x)=
\ep(x)\Big||\det x|^{1\over n}\Big|=\cases{
\tau\hbox{ for }n=1\cr  
\ep(x_0)\vth(x^2)\sqrt{x^2}\hbox{ for }n=2\cr}
\end{eqn}In general for $n\ge2$, it is not linear
$\tau(x+y)\ne \tau(x)+\tau(y)$.

\subsection{Timespace Manifolds}

The  complex  algebra $\C(n)$ with the commutator as Lie bracket 
is, on the one side as complex $n^2$-di\-men\-sio\-nal
space, the rank $n$ Lie algebra
\footnote{\scriptsize
The Lie algebra of the Lie group $G$ is denoted by $\log G$.}
 of the complex Lie group $\GL(\C^n)\subnoteq\C(n)$ and, on the other side
as real $2n^2$-di\-men\-sio\-nal space, the rank $2n$ Lie algebra of the
real group\footnote{\scriptsize
$\C_\R$ denotes the real 2-di\-men\-sio\-nal structure 
$\R\pl i\R$ of the complex numbers $\C$.}
 $\GL(\C^n_\R)$.
The antisymmetric space $i\R(n)$ in $\C(n)$ is the rank $n$ 
Lie algebra of  the unitary group $\U(n)$ 
\begin{eqn}{rl}
\C(n)=\log\GL(\C^n),~~&\GL(\C^n)=\exp\C(n)\cr
\R(n)\pl i\R(n)=\log \GL(\C^n_\R),~~&
\GL(\C^n_\R)=\exp\brack{\R(n)\pl i\R(n)}\cr
i\R(n)=\log\U(n),~~&\U(n)=\exp i\R(n)\cr 
\end{eqn}

The vector space of the ti\-me\-spa\-ce translations $\R(n)$ is isomorphic 
to the quotient of the full 
with respect to the unitary Lie algebra. Its exponent 
is isomorphic to the corresponding ho\-mo\-ge\-neous space,
the  real $n^2$-di\-men\-sio\-nal manifold 
with the orbits $g\U(n)$ of the unitary group $\U(n)$ in the full group 
$\GL(\C^n_\R)$
\begin{eqn}{l}
\R(n)\cong\log\GL(\C^n_\R)/\log\U(n),~~\D(n)=\exp \R(n)\cong \GL(\C^n_\R)/\U(n)
\end{eqn}The 'compact in complex manifold' $\D(n)$ will be called 
a {\it ti\-me\-spa\-ce manifold}, it is isomorphic to the 
strictly positive timelike translations
\begin{eqn}{l}
\D(n)=\{d= \ro e~\ro e^\star\mid\ro e\in\GL(\C^n_\R)\}\cong\R(n)^+_\Time
\end{eqn}i.e. the ti\-me\-spa\-ce manifold 
$\D(n)$ can be embedded into its tangent space $\R(n)$
where light and space translations arise as  genuine tangent phenomena.  

Timespace  $\D(n)$  is totally semiordered $\sqsubseteq$
(transitive and reflexive)
via the abelian projection onto the totally ordered group $\D(1)$
\begin{eqn}{l}
\det:\D(n)\map\D(1),~~d\mape \det d\cr
d_{1,2}\in\D(n):~~d_1\sqsubseteq d_2\iff \det d_1\le\det d_2\cr
\end{eqn}Therewith, $\D(n)$ will be  also called a 
{\it causal manifold} \cite{HIL}.
The set of all semiorder induced equivalence classes
is isomorphic to the causal group $\D(1)$ and carries a total order. 

Also the name {\it modulus manifold} is justified
for $\D(n)$ since it parametrizes
the possible scalar products of the complex space $\C^n$ 
with basis $\{e^A\}_{A=1}^n$ where
the C$^*$-algebra $\C(n)$ is acting on
\begin{eqn}{rl}
\D(n)\ni
d= \ro e~\ro e^\star\cong {\scriptsize\pmatrix{
\sprod{e^1}{e^1}&\dots&\sprod{e^1}{e^n}\cr
\cdots&\cdots&\cdots\cr
\sprod{e^n}{e^1}&\dots&\sprod{e^n}{e^n}\cr}}
\end{eqn}

The real Lie groups\footnote{\scriptsize
$\U(1_n)\cong\U(1)$ and $\D(1_n)\cong\D(1)$ 
denote the scalar phase and causal (dilatation) group in $\GL(\C^n_\R)$. 
$\bl 1(n)$
is the unit of the $\C^n$-automorphism groups.}
for the ti\-me\-spa\-ce manifold 
involve as factors the 
{\it causal group} $\D(1)$
(direct factor, denoted by $\x$), 
the {\it phase group} $\U(1)$ and the {\it special groups}
$\SL(\C^n_\R)$ and $\SU(n)$ resp.
They will be called
\begin{eqn}{lrl}
\hbox{\it external group: }
&\GL(\C^n_\R)&=\D(1_n)\x \UL(n)\cr
\hbox{\it internal group: }
&\U(n)&=\U(1_n)\o\SU(n)\cr
\hbox{\it ti\-me\-spa\-ce manifold: }
&\GL(\C^n_\R)/\U(n)&\cong \D(1)\x\bl{SD}(n)\cr
\end{eqn}The 2nd direct factor 
$\bl {SD}(n)=\SL(\C^n_\R)/\SU(n)$ in the manifold
will be called the {\it Sylvester or boost submanifold}.
It is trivial only for the abelian case $n=1$.

The unimodular\footnote{\scriptsize
Some authors use 'unimodular' (here defined by $|\det g|=1$) for 'special'
(here defined by  $\det g=1$).}
 groups $\UL(n)=\U(1_n)\o\SL(\C^n_\R)$
(external)  and $\U(n)$ (internal)  have with the cyclotomic group
$\I_n=\{\al\in\C\mid\al^n=1\}$ the centers, 
phase correlation groups and adjoint groups\footnote{\scriptsize
The adjoint group of a group is defined as the classes up to the center
$\Ad G=G/\centr G$.}
\begin{eqn}{rl}
\centr\UL(n)\cong\U(1),~~&\centr\U(n)\cong\U(1)\cr
\U(1_n)\cap \UL(n)\cong\I_n,~~&\U(1_n)\cap\SU(n)\cong\I_n\cr~
\Ad\UL(n)\cong\SL(\C^n_\R)/\I_n,~~&\Ad\U(n)\cong\SU(n)/\I_n\cr
\end{eqn}

As a homogeneous space, 
a ti\-me\-spa\-ce manifold has a nontrivial external action 
(from left on $g\U(n)\in\D(n)$) with the causal group 
and the adjoint external group
\begin{eqn}{l}
\hbox{external action on $\D(n)$ with } \D(1)\x {\SL(\C^n_\R)/\I_n}
\end{eqn}and a trivial action (from right) with the internal group
(therefore the name 'internal').

There is another more familiar 
chain of causal ti\-me\-spa\-ce manifolds, characterized by
the orthogonal real structures 
\begin{eqn}{l}
s=0:~~\D(1)  ,~~
s\ge1:~~\D(1)\x\SO^+(1,s)/\SO(s)
\end{eqn}with $s\ge0$ space
dimensions. This chain
meets the $\GL(\C^n_\R)/\U(n)$ chain only for
the two ti\-me\-spa\-ce dimensions $n^2=1+s=1,4$.
Obviously, the orthogonal structures involve
an invariant  bilinear form for all dimensions. 

For $n=2$ one has the isomorphy 
with the or\-tho\-chro\-nous Lorentz group $\SO^+(1,3)$ and the rotation group
$\SO(3)$
\begin{eqn}{rll}
\UL(2)/\U(1)&\cong \SL(\C^2_\R)/\I_2&\cong \SO^+(1,3)\cr
\U(2)/\U(1)&\cong \SU(2)/\I_2&\cong \SO(3)\cr
\D(2)&=\GL(\C^2_\R)/\U(2)&\cong\D(1)\x\SO^+(1,3)/\SO(3)\cr
\end{eqn}If one visualizes the 
real 4-di\-men\-sio\-nal ti\-me\-spa\-ce manifold $\D(2)$ 
embedded as the strict future
cone $\R(2)_\Time^+$ in the Min\-kow\-ski translations $\R(2)$, 
this cone has to be foliated\footnote{\scriptsize
Take the 3-di\-men\-sio\-nal projection with hyperboloids $\SO^+(1,2)/\SO(2)$
with the 2-di\-men\-sio\-nal tangent planes.}
 with the  
hyperboloids $\SO^+(1,3)/\SO(3)$
(hyperbolic foliation). The causal group $\D(1)$ action
on the manifold connects different hyperboloids by 'hyperbolic hopping'
whereas the or\-tho\-chro\-nous group $\SO^+(1,1)$ action 
on the individual hyperboloids can be described as 'hyperbolic stretching'. 
The Min\-kow\-ski translations $\R(2)$ as tangent structure of the ti\-me\-spa\-ce
manifold $\D(2)$ can be 
visualized 
with a 3-di\-men\-sio\-nal tangent plane of a timelike hyperboloid 
$\SO^+(1,3)/\SO(3)$ and the 
tangent line of 'blowing up' or 'shrinking' this hyperboloid with $\D(1)$.  
The causal order is the 'foliation order' of 
the positive timelike hyperboloids.

\subsection{The Rank of Timespace}

A ti\-me\-spa\-ce translation  is diagonalizable with a 
unitary matrix
\begin{eqn}{rl}
x=x^\star\in\R(n):~~&x=u(x)\o \diag(x)\o u(x)^\star\cr
\end{eqn}The relativistic case $n=2$ 
uses in addition to two Cartan coordinates 
two polar coordinates $(\phi,\th)$ from the unit sphere
$\SO(3)/\SO(2)$
\begin{eqn}{rl}
n=2:&
{1\over2}{\scriptsize\pmatrix{
x_0+x_3&x_1-ix_2\cr x_1+ix_2&x_0-x_3\cr}}
=u(x)
{\scriptsize\pmatrix{
{x_0+|\rvec x|\over2}&0\cr 0&{x_0-|\rvec x|\over2}\cr}}
u(x)^\star\cr
&~~\cr
&u(x)={\scriptsize\pmatrix{
\cos{\th\over2}&-e^{-i\phi}\sin{\th\over2}\cr
e^{i\phi}\sin{\th\over2}&\cos{\th\over2}\cr}}\hbox{ for }
\rvec x=|\rvec x|{\scriptsize\pmatrix{
\cos\phi\sin\th\cr
\sin\phi\sin\th\cr
\cos\th\cr}}\cr
\end{eqn}

Timespace  translations and ti\-me\-spa\-ce  manifold are isomorphic
as real manifolds
\begin{eqn}{l}
\exp:\R(n)\map \D(n),~~x\mape e^x\cr
\log:\D(n)\map \R(n),~~d\mape \log d \cr
\end{eqn}using the  exponentation and logarithm of the
diagonal matrices
\begin{eqn}{l}
x=u(x)\o \diag(x)\o u(x)^\star\then e^x=u(x)\o e^{\diag(x)}\o u(x)^\star\cr
d=u(d)\o \diag(d)\o u(d)^\star\then \log d=u(d)\o \log \diag(d)\o u(d)^\star\cr
\end{eqn}e.g. for $n=2$
\begin{eqn}{rl}
n=2:e^{{x_0\bl 1(2)+\rvec x\rvec \si\over2}}&=
e^{{x_0\over2}}(\bl 1(2) \cosh{|\rvec x|\over2}+
{\rvec\si |\rvec x|
\over|\rvec x|}\sinh{|\rvec x|\over2})\cr
&=e^{{x_0\over2}}{\scriptsize\pmatrix{
\cosh{|\rvec x|\over2}+\cos\th\sinh{|\rvec x|\over2}
&e^{-i\phi}\sin\th\sinh{|\rvec x|\over2}\cr
e^{i\phi}\sin\th\sinh{|\rvec x|\over2}
&\cosh{|\rvec x|\over2}-\cos\th\sinh{|\rvec x|\over2}\cr}}\cr 
&=u(x){\scriptsize\pmatrix{
e^{{x_0+|\rvec x|\over2}}&0\cr 0&e^{{{x_0-|\rvec x|\over2}}}\cr}}
u(x)^\star\cr
\end{eqn}For $n=1$, one has an isomorphy for the abelian groups
 $\D(1)=e^\R\cong\R$.

In the general case,
the diagonal matrix for a ti\-me\-spa\-ce point 
contains the $n$ strictly positive
spectral values
\begin{eqn}{l}
d\in\D(n):~~
\diag(d)=
{\scriptsize\pmatrix{
e^{\xi_1}&0&\dots&0\cr
0&e^{\xi_2}&\dots&0\cr
\dots&&\dots&\cr
0&0&\dots&e^{\xi_n}\cr}},~~x_0={\SUM_{r=1}^n}\xi_r
\end{eqn}which realize $n$ times the abelian 
causal group $\D(1)$. 
In a nontrivial
boost submanifold $\bl{SD}(n)$, the group $\D(1)$ comes in the selfdual
decomposable re\-pre\-sen\-ta\-tion ('Prokrustes re\-pre\-sen\-ta\-tion'),
isomorphic  to the
or\-tho\-chro\-nous group $\SO^+(1,1)$ 
\begin{eqn}{l}
\D(1)\cong \SO^+(1,1)\ni d(\xi)={\scriptsize\pmatrix{
\cosh{\xi\over2}&\sinh{\xi\over2}\cr
\sinh{\xi\over2}&\cosh{\xi\over2}\cr}}\cong
{\scriptsize\pmatrix{
e^{\xi\over2}&0\cr 0&e^{-{\xi\over2}}\cr}}\cr
(-{d^2\over d\xi^2}+1)d(\xi)=0
\end{eqn}

The unitary 
diagonalization transformation  is determined up to the 
diagonal phases
\begin{eqn}{l}
u(d), u(x)\in \SU(n)/\U(1)^{n-1}
\end{eqn}

Therewith,  ti\-me\-spa\-ce is  isomorphic - as manifold - to a 
product of noncompact causal groups 
(Cartan subgroup) and a compact manifold
\begin{eqn}{rl}
\GL(\C^n_\R)/\U(n)=\D(n)&\cong \D(1)^n\x\SU(n)/\U(1)^{n-1}\cr
\SL(\C^n_\R)/\SU(n)=\bl{SD}(n)&\cong \D(1)^{n-1}\x\SU(n)/\U(1)^{n-1}\cr
\end{eqn}and for the ti\-me\-spa\-ce translations
\begin{eqn}{l}
\log\GL(\C^n_\R)/\log\U(n)\cong \R(n)\cong\R^n\x\SU(n)/\U(1)^{n-1}\cr
\end{eqn}The abelian group $\D(1)^n$ and 
its Lie algebra $\R^n$ constitute  the Cartan 
skeleton of the ti\-me\-spa\-ce manifold $\D(n)$ and 
the ti\-me\-spa\-ce translations $\R(n)$ resp.

The rank \cite{HEL} 
$n$ and $n-1$ 
of the  homogeneous manifold $\D(n)$ and  $\bl{SD}(n)$ resp.
has to be seen in analogy to the rank of a Lie algebra or its Lie group,
e.g. rank $n$ and  $n-1$  for $\U(n)$ and $\SU(n)$ resp. 
with the manifold factorizations
 \begin{eqn}{rl}
\U(n)&\cong\U(1)^n\x \SU(n)/\U(1)^{n-1}\cr
\SU(n)&\cong\U(1)^{n-1}\x \SU(n)/\U(1)^{n-1}\cr
\end{eqn}

\subsection{Tangent Structure and Poin\-ca\-r\'e Group}

A  ti\-me\-spa\-ce $\D(n)$ and $\R(n)$ analysis 
is interpreted and performed with 
the linear forms 
(dual space\footnote{\scriptsize
$V^T$ denotes the dual vector space (linear forms) for a vector space $V$
with the bilinear dual product
$V^T\x V\map\C$, $\dprod\om v=\om(v)$.})
$\R(n)^T$ of the ti\-me\-spa\-ce translations 
containing  the weights (collection of eigenvalues). 
The linear forms  will be called
the {\it frequency (energy)} space for $n=1$ and the 
{\it energy-momenta} space for
 $n\ge2$. In the re\-pre\-sen\-ta\-tion by the matrix algebra $\C(n)$
the 'double trace 
with one open slot'
describes an isomorphism  between translations and energy-momenta
 \begin{eqn}{l}
 \R(n)\map\R(n)^T,~~q\mape \d q =\tr q\o...\cr
\hbox{dual product: } 
 \R(n)^T\x\R(n)\map\R,~~\dprod{\d q} x=\tr q\o x\cr
 \end{eqn}With generalized Pauli matrices one has as dual bases
 \begin{eqn}{rl} 
 \hbox{$\R(n)$-translations basis:}&\{\rho(n)^j\}_{j=0}^{n^2-1}
 =\{{\bl 1(n)\over n},{\si(n)^a\over2}\}_{a=1}^{n^2-1}\cr
\hbox{$\R(n)^T$-energy-momenta basis:}&\{\d\rho(n)_j\}_{j=0}^{n^2-1}
 =\{\bl 1(n),\si(n)^a\}_{a=1}^{n^2-1}\cr
\hbox{dual bases:}&\tr\rho(n)^j\o\d\rho(n)_k=\de^j_k\cr
\end{eqn} 

The Cartan coordinates $\{\xi_r\}_{r=1}^n$  for the
translations have their correspondence in
{\it Cartan mas\-ses} $\{\mu_r\}_{r=1}^n$
for the energy-momenta, positive for positive
energy-momenta  
\begin{eqn}{rl}
q\in\R(n)^T:~~&q= u(q){\scriptsize\pmatrix{
\mu_1&0&\dots&0\cr
0&\mu_2&\dots&0\cr
\dots&& \dots&\cr
0&0&\dots&\mu_n\cr}}u(q)^\star\cr
&q\succeq 0\iff\hbox{ all }\mu_r\ge0\cr
\end{eqn}

The   external group $\GL(\C^n_\R)$ action
on the causal manifold $\D(n)$ induces a faithful 
action of the
adjoint external group $\Ad \GL(\C^n_\R)\cong\SL(\C^n_\R)/\I_n$ on the
tangent structures. This defines the
{\it Poin\-ca\-r\'e group} as semidirect product 
(symbol $\x_s$)    of the adjoint 
external group
and the
additive vector space  groups
\begin{eqn}{l}
s\in\SL(\C^n_\R):~~\left\{\begin{array}{ll}
\Ad_\star s:\R(n)\map\R(n),&\Ad_\star s(x)=s\o x\o s^\star\cr
\Ad_\star \hat s:\R(n)^T\map\R(n)^T,&\Ad_\star \hat s(q)
= \hat s\o q\o \hat s^\star\cr
&\hbox{with }\hat s=s^{-1\star}\end{array}\right.\cr
\bl{POIN}(n)=\SL(\C^n_\R)/\I_n\x_s \R(n) 
\end{eqn}

In contrast to the direct product structure $\D(1)\x\bl{SD}(n)$ of the 
causal manifold,
a decomposition 
of the ti\-me\-spa\-ce translations into time translations $\T$ and space
translations $\S(n-1)$ or 
of the dual space into energy and momenta  is
incompatible with the action of $\SL(\C^n)/\I_n$ for $n\ge2$, it is only
compatible with the action of the adjoint compact 
subgroup  $\SU(n)/\I_n$.

Two remarks are in order: In general,
the adjoint action of a Lie group $G$
on its Lie algebra $\log G$ can be  characterized by 
$l\mape \Ad g(l)=glg^{-1}$ leading to the
semidirect product $\Ad G\x_s T(G)$.
The abelian  
normal subgroup $T(G)$ is the vector space structure of the Lie algebra 
$\log G\cong T(G)$, 
i.e. a nonabelian 
Lie bracket has to be 'forgotten' \cite{LIE13} for the translations $T(G)$.  

In the case 
of a group with conjugation $*$, 
the involution $g\mape\hat g=g^{-1*}$
is a group  automorphism with the  
$*$-unitary group 
$U(G,*)=\{u\in G\mid u^*=u^{-1}\}$ as invariants.
Correspondingly, the Lie algebra $\log G$ of a complex
 finite dimensional Lie group with conjugation $*$ 
is the
direct sum of the isomorphic antisymmetric and symmetric 
real vector spaces $\log G_\pm$ with $l^*_\pm=\pm l_\pm$.
$\log G_-$ is the real Lie algebra of the unitary group $U(G,*)$.
In order to be compatible 
with the conjugation $g\mape g^*$ and $l\mape l^*$,
the adjoint action has to
be modified to $l\mape \Ad_*g(l)=glg^*$. 
For the $*$-unitary subg $U(G,*)$, one has $\Ad u=\Ad_*u$.
Both subspaces $\log G_\pm$ remain stable. The emerging semidirect
adjoint group is $\Ad_* G\x_s\log G_+$ with the additive 
group structure of the vector space $\log G_+\cong \log G/\log U(G,*)$.
For all adjoint actions, the abelian center of the group is represented 
in the semidirect product  via
the additive structure of the  translation factor.

For time alone, $n=1$, the Poin\-ca\-r\'e  group is the additive structure
of the time translations 
\begin{eqn}{l}
\bl{POIN}(1)=\R(1)=\log\D(1)
\end{eqn}

In the familiar relativistic case, $n=2$, with
\begin{eqn}{l}
\bl{POIN}(2)\cong\SO^+(1,3)\x_s\R(2)\cr
\R(2)\cong\log\D(1)\pl\log\SO^+(1,3)/\log \SO(3)\cr
\end{eqn}the Min\-kow\-ski time-, light- and spacelike translations are 
isomorphic to  ho\-mo\-ge\-neous manifolds with characteristic fixgroups
('little' groups) for the translations
\begin{eqn}{lll}
\R(2)_\Time^\pm&\cong\D(1)\x\SO^+(1,3)/\SO(3)&\cong
\GL(\C^2_\R)/\U(2)\cr
\R(2)_\Light^\pm&\cong\SO^+(1,3)/\SO(2)\x_s\R^2&
\cong\SL(\C^2_\R)/\U(1)\x_s\C_\R\cr
\R(2)_\Space\setminus\{0\}&\cong\D(1)\x\SO^+(1,3)/\SO^+(1,2)&\cong
\GL(\C^2_\R)/\U(1,1)\cr
\end{eqn}The semidirect product $\SO(2)\x_s\R^2$ 
fixgroup of the lightlike translations is the Euclidean 
group.\footnote{\scriptsize
For the Poin\-ca\-r\'e  group $\SO^+(1,s)\x_s\R^{1+s}$ 
with $s\ge1 $ space dimensions the corresponding fixgroups are
$\SO(s)$ (timelike)
, $\SO(s-1)\x_s\R^{s-1}$ (lightlike) and $\SO^+(1,s-1)$ 
(spacelike). 
The lightlike translations
fixgroup for $s=1$ is trivial $\{1\}$.}

For the case $n\ge2$, the disjoint decompositions
of the spacelike and timelike manifolds  
into $(n-1)$ submanifods (subsection 2.1) reflect different 
unitary fixgroups
\begin{eqn}{rl}
n\ge2:~~\left\{\begin{array}{rl}
\R(m,n-m)_\Space&\cong\GL(\C^n_\R)/\U(m,n-m)\cr
\R(m,n-1-m)^\pm_\Light&\cong\SL(\C^n_\R)/\U(m,n-1-m)\x_s\C_\R^{n-1}\cr
&m=1,\dots,n-1\end{array}\right.
\end{eqn}

The  fixgroup $\U(n)$ for the action of the 
external group  $\SL(\C^n_\R)/\I_n$
on the strictly positive translations $\R(n)^+_\Time$ 
should not be confused with the internal group 
$\U(n)$ which acts trivially (from right)  on ti\-me\-spa\-ce
$\GL(\C^n_\R)/\U(n)$. A group $G$ acting from left on
the subgroup classes $gU\in G/U$ has an $U$-isomorphic fixgroup
for any point $gU$. 

\section{Representations for Timespace}

The solution or an experimentally oriented formulation of
a ti\-me\-spa\-ce dynamics  requires an
analysis, e.g. of a symmetry invariant,  with  respect 
to the operations used in the definition
of a ti\-me\-spa\-ce manifold $\D(n)=\GL(\C^n_\R)/\U(n)$.
Starting from a purely algebraic framework, one can even say
that an ana\-ly\-sis with respect to
time or ti\-me\-spa\-ce  re\-pre\-sen\-ta\-tions {\it introduces} 
the time or the ti\-me\-spa\-ce dependence
of quantum mechanical operators \cite{WEIZWP} or relativistic fields.

For a solution of a dynamics, the involved 
nondecomposable re\-pre\-sen\-ta\-tions  of the external-internal 
real Lie group
\begin{eqn}{l}
\GL(\C^n_\R)\x \U(n)
\end{eqn}have to be determined.
The eigenvalue and eigenvector problems to be solved in a quantum structure is 
classically expressed by equations of motion.
The eigenvalues (weights) 
for the action of a
real Lie group are linear forms of the Cartan subalgebra and, therefore, 
have to be real.
All weights (collection of eigenvalues) form a subgroup, discrete or
continuous, in the additive group of the linear forms on the
Lie algebra of the external-internal group.

The theorem that two diagonalizable finite dimensional endomorphisms $f,g$ are
simultaneously diagonalizable, i.e. have a common eigenvector basis,
if, and only if, they commute with each other $\com fg=0$, is the mathematical
formalization of a central
 quantum operational structure. However, operators cannot be identified with
 states, mathematically: In general, endomorphisms $f$ of a complex 
 finite dimensional space
allow only a Jordan triangularisation, i.e. no eigenvector basis.
In general, the {\it nondecomposable} re\-pre\-sen\-ta\-tion spaces 
of the external-internal group can be 
spanned by principal vectors.
{\it Irreducible} re\-pre\-sen\-ta\-tions are a special case,
their vector spaces can be spanned even by eigenvectors   
(diagonalization in the semisimple case). 

For $n=1$ with time as the  causal group $\D(1)$,
one has a  dynamics for mass points (mechanics) where a
Hamiltonian  $H$ representing - or defining - 
the time translations $\R(1)$ as
generator of the causal group
is analyzed with respect to 
the  involved re\-pre\-sen\-ta\-tions of time $\D(1)$, illustrated 
by the equation of motion ${d a\over d t}=\com {iH}a$
for a quantum operator $a$, e.g. position $X$ or momentum $P$. 
E.g.  the Hamiltonian for the $N$-di\-men\-sio\-nal 
isotropic harmonic oscillator 
$H={P^2+X^2\over2}$ as $\SU(N)$-invariant or the hydrogen Hamiltonian
$H={\rvec P^2\over2}-{1\over|\rvec X|}$  as
an invariant of the group ${\SU(2)\x\SU(2)\over\I_2}\cong\SO(4)$ 
(elliptic bound states)
or $\SL(\C^2_\R)/\I_2\cong\SO^+(1,3)$ 
(hyperbolic scattering states), both subsymmetries of
$\SU(2,2)/\I_2\cong\SO(2,4)$. The time dependence of 
quantum operators  can be
introduced (defined), e.g. by $a(t)=e^{iHt}a e^{-iHt}$ 

For relativistic fields, $n=2$, with the  ho\-mo\-ge\-neous ti\-me\-spa\-ce 
$\D(2)=\GL(\C^2_\R)/\U(2)$ and the 
tangent Min\-kow\-ski translations $\R(2)$,
a dynamics analyzes an interaction 
with respect to  the external-internal group,
e.g. the analysis of invariant  gauge vertices 
like $\ol{\bl\Psi}\ga_k\bl A^k\bl\Psi$ 
with fermion and gauge degrees of freedom  $\bl\Psi$ and
$\bl A$ resp. in
the standard model of elementary particles. 
Involving  both noncompact and nonabelian structures, 
this much more complicated analysis 
works with re\-pre\-sen\-ta\-tions of the 
external noncompact causality $\D(1)$ and Lorentz 
$\SL(\C^2_\R)$ properties as well as internal
compact hypercharge $\U(1)$ and isospin $\SU(2)$ properties, 
as expressed by field equations,
e.g. $\ga_k{\p\bl\Psi\over\p x_k}
=i\cl H(\bl\Psi)=i\ga_k\bl A^k\bl\Psi$.

\subsection{Representations of the Causal Group}

Any dynamics requires a {\it causal  analysis} with respect
to the group $\D(1)=\{e^\tau\mid\tau\in\R\}$.

All $\D(1)$ re\-pre\-sen\-ta\-tions 
 \cite{BOE,S89} can be built from nondecomposable  re\-pre\-sen\-ta\-tions. The
finite di\-men\-sio\-nal, nondecomposable, unitary 
 complex re\-pre\-sen\-ta\-tions 
$ e^\tau\mape  ( N|m)(\tau)$ of
the real Lie group $\D (1)$
are characterized by positive integers $ N\in\N$ 
for the dimension $ 1+N $ 
of the re\-pre\-sen\-ta\-tion space and 
a real  number $m$ - a
frequency (energy)  for time  and a mass for ti\-me\-spa\-ce  -
from the dual group,
the linear frequency (mass)  space $\log\D(1)^T\cong\R$.
They involve a power $ 1+N $ {\it nilpotent} part $\La_N$
 (nil-Hamiltonian \cite{S89}), nontrivial for
$N\ne0$ 
\begin{eqn}{l}
\D(1)\ni e^\tau\mape   (N|m)(\tau)
=e^{i\La_N \tau}e^{i m \tau}\hbox{ with }\cases{
 N=0,1,2,\dots\cr
(\La_N)^{ N}\ne0,~N\ne0\cr
(\La_N)^{ 1+N }=0\cr}\cr
   ( N|m)(\tau)=e^{\La_N{d\over d m}}   e^{im\tau},~~
   \tr( N|m)(\tau)=( 1+N ) e^{im\tau}\cr   
\end{eqn}The  following explicit examples 
with nilcyclic matrices $\La_N$ illustrate the abstract structure
\begin{eqn}{rl}
   (0|m)(\tau)&= e^{im \tau} \cr
      (1|m)(\tau)&\cong {\scriptsize\pmatrix{
1&i \la\tau\cr 0&1\cr}}e^{im\tau }=
{\scriptsize\pmatrix{
1&\la{d\over  d m}\cr 0&1\cr}} e^{im\tau}\cr
~~\cr
   (2|m)(\tau)&\cong {\scriptsize\pmatrix{
1&i \la\tau&{(i\la\tau)^2\over 2!}\cr 
0&1&i\la\tau\cr
0&0&1\cr}}e^{im\tau }=
{\scriptsize\pmatrix{
1&\la{d\over  d m}&{\la^2\over2!}{d^2\over  d m^2}\cr
 0&1&\la{d\over  d m}\cr
 0&0&1}}  e^{im\tau}\cr
\La_0=0,&~
\La_1\cong
\la{\scriptsize\pmatrix{
0&1\cr 0&0\cr}}
,~~\La_2\cong
\la{\scriptsize\pmatrix{
0&1&0\cr 0&0&1\cr 0&0&0\cr}}\cr
  \end{eqn}

To take care of the {\it real} structure of the causal group $\D(1)$, 
a {\it complex} $\D(1)$-re\-pre\-sen\-ta\-tion has to be a 1-di\-men\-sio\-nal 
subgroup of 
a {\it unitary group}. Therewith the constant $ m$ has to be real in
nondecomposable re\-pre\-sen\-ta\-tions.
The $\D(1)$-images for the 1-di\-men\-sio\-nal  
re\-pre\-sen\-ta\-tions $  (0| m)$ for $ m\ne0$
(harmonic oscillator)  are isomorphic
to $\U(1)$. The re\-pre\-sen\-ta\-tions $  (1| m)$ 
(e.g. a free mass point for $ m=0$) have real 1-di\-men\-sio\-nal faithful images 
in the {\it indefinite  
unitary} group $\U(1,1)$, $  ( 2| m)$ in $\U(2,1)$, 
$(3| m) $ in $\U(2,2)$ etc.
\begin{eqn}{l}
(N|m)\hbox{ in }\U(N_+,N_-)\hbox{ with}
\cases{
N_++N_-= 1+N \cr
N_+-N_-=\cases{
1,~~ N=0,2,\dots\cr
0,~~ N=1,3,\dots\cr}\cr}
\end{eqn}

In addition to the discrete 
dimension $ 1+N $, the nondecomposable 
causal re\-pre\-sen\-ta\-tions involve
 two continuous constants, $m$ and $\la$:
Only the frequency (mass) $ m$ 
is an invariant of the 
causal group $\D(1)$. It is the {\it causal unit
of the re\-pre\-sen\-ta\-tion}.
The matrix form of the nilpotent $\La_N$ 
and the real constant\footnote{\scriptsize
E.g. for $ N=1$ with $\La_1=\la{\scriptsize\pmatrix{
0&1\cr 0&0\cr}}$
 the
constant $\la$ is transformed with
 $g={\scriptsize\pmatrix{e^{\al}&0\cr0&e^{-\al}\cr}}$ 
to $\la\mape e^{2\al}\la$. A similar structure
 \cite{BRS,S911} is used
for the 'gauge fixing constant' in quantum gauge theories. 
The gauge fixing constant  has to be nontrivial - 
its value is physically irrelevant. The gauge structure 
in the BRS formulation is nilpotent.}
$\la\ne0$ are determined up to
equivalence $g\La_N g^{-1}$.

Re\-pre\-sen\-ta\-tions with opposite frequency (mass) $(N|\pm m)$ 
are dual to each other and nonequivalent for $m^2>0$.
The selfdual, $m^2$-dependent  formulation contains 
two dual re\-pre\-sen\-ta\-tions, e.g. in $\SO(2)$ for $N=0$
\begin{eqn}{l}
  (0|m)(\tau )\pl (0|-m)(\tau)=( 0||m^2)(\tau)\cong
{\scriptsize\pmatrix{e^{im\tau}&0\cr 0&e^{-im\tau}\cr}}
\cong 
{\scriptsize\pmatrix{
\cos m\tau &{i\over m  }\sin m\tau  \cr 
im  \sin m\tau  &\cos m\tau \cr}} \cr
\end{eqn}

Only the positive unitary 1-di\-men\-sio\-nal re\-pre\-sen\-ta\-tions 
$(0|m)$ are irreducible, they are
unfaithful - the $\U(1)$-represented time is periodic.
The indefinite unitary faithful re\-pre\-sen\-ta\-tions $(N|m)$ for 
$ N\ge1$ are 
reducible, but nondecomposable. 
The lowest di\-men\-sio\-nal 
faithful $\D(1)$ re\-pre\-sen\-ta\-tions $(1|m)$ will be called {\it fundamental}. 

The   pairs $(N|m)$ with a positive integer (dimension) and 
a real number (causal unit) 
form the  abelian  {\it representation monoid of the causal group}
for all equivalence classes 
of  {\it nondecomposable} causal re\-pre\-sen\-ta\-tions
\begin{eqn}{rl}
\ro{mon}~\D(1)
=&\{(N|m)\mid N=0,1,\dots,~~m\in\R\}=\N\x\R\cr
&( N_1|m_1)+( N_2|m_2)=( N_1+ N_2|m_1+m_2)\cr
&\hbox{neutral element: }(0|0)\cr
\end{eqn}The {\it weight group} of $\D(1)$ is the regular  subgroup of 
the representation monoid, it 
characterizes the {\it irreducible} re\-pre\-sen\-ta\-tions of the causal 
group and is the dual space $\log\D(1)^T$ of the causal translations
$\log\D(1)$
\begin{eqn}{rl}
\ro{grp}~\D(1)&=\{(0|m)\mid m\in\R\}=\R\cr
\ro{mon}~\D(1)&\supnoteq\ro{grp}~\D(1)
\end{eqn}

The {\it $\U(1)$-weights} (oriented winding numbers) 
for the compact quotient group $\U(1)\cong\D(1)/e^\Z$
form a discrete subgroup, the {\it representation group for $\U(1)$}
\begin{eqn}{rl}
\ro{grp}~\U(1)=\{(0|Z)\mid Z\in\Z\}=\Z\cr
\end{eqn}

\subsection{Representations of $\GL(\C_\R)$}
 
 All  {\it irreducible} complex re\-pre\-sen\-ta\-tion of
the real abelian Lie group
$\GL(\C_\R)
=\D(1)\x\U(1)$
are 1-di\-men\-sio\-nal and characterized by an integer winding number
($\U(1)$-weight)  and a complex number 
\begin{eqn}{l}
\GL(\C_\R)\ni\de=e^{\tau+i\al}\mape
|\de|^{i m}({\de\over\ol\de})^{{Z\over 2}}=\de^{{i m+Z\over2}}
\ol \de^{{i m-Z\over2}}
=e^{i m\tau} e^{Zi\al}\cr
(m;Z)\in\C\x\Z
\end{eqn}

Complex re\-pre\-sen\-ta\-tions of real groups have to be
unitary. Unitary  irreducible 
$\D(1)$-re\-pre\-sen\-ta\-tions are necessarily positive unitary.
All   unitary irreducible re\-pre\-sen\-ta\-tions have a
real causal unit $ m$ ($\D(1)$-weight)
\begin{eqn}{rl}
\ro{grp}~\GL(\C_\R)
&=\ro{grp}~\D(1)\x\ro{grp}~\U(1)=\{(m;Z)\}= \R\x\Z
\end{eqn}As seen in the former subsection, the group of the 
 {\it $\GL(\C_\R)$-weights}
is the regular group of the
{\it $\GL(\C_\R)$-representation monoid} for all equivalence classes of the 
unitary {\it nondecomposable} re\-pre\-sen\-ta\-tions
\begin{eqn}{rl}
\ro{mon}~\GL(\C_\R)
&=\ro{mon}~\D(1)\x\ro{grp}~\U(1)=\{(N|m;Z)\}\cr
&=\N\x \R\x\Z\cr
\ro{mon}~\GL(\C_\R)&\supnoteq \ro{grp}~\GL(\C_\R)
\end{eqn}

\subsection{Finite Dimensional Representations of $\SL(\C^n_\R)$}

All complex  re\-pre\-sen\-ta\-tions of 
the compact real  $(n^2-1)$-di\-men\-sio\-nal Lie group
$\SU(n)$, $n\ge2$, are decomposable into irreducible ones
which have finite dimensions. 
The irreducible  $\SU(n)$-re\-pre\-sen\-ta\-tions
 are  characterized 
 - according to rank $(n-1)$ and Cartan subgroup $\U(1)^{n-1}$ - 
 by $(n-1)$ positive integers with the additive  
 {\it $\SU(n)$-representation monoid}
for the equivalence classes
\begin{eqn}{l}
\ro{mon}~\SU(n)=\{
\brack{2J_1,\dots,2J_{n-1}}\mid 2J_k=0,1,\dots\}=\N^{n-1}
\end{eqn}The positive integers reflect the $\N$-linear 
combination of the $(n-1)$ {\it fundamental} re\-pre\-sen\-ta\-tions
$\brack{1,0,\dots,0}$, ..., $\brack{0,\dots,0,1}$,
e.g. the Pauli spinor re\-pre\-sen\-ta\-tion $\brack1$ for  $\SU(2)$
or
the quark and antiquark representations 
$\brack{1,0}$ and $\brack{0,1}$ for $\SU(3)$.
The adjoint re\-pre\-sen\-ta\-tion $\brack{1,0,\dots,0,1}$
is faithful only for the adjoint group $\SU(n)/\I_n$, e.g. the adjoint
$\SU(2)$-re\-pre\-sen\-ta\-tion $\brack 2$ for $\SO(3)$ or
the $\SU(3)$-octet representation $\brack{1,1}$ for $\SU(3)/\I_3$.

The $\SU(n)$-representation monoid is 
the positive cone (dominant weights) 
in all {\it $\SU(n)$-weights} which 
constitute  a discrete subgroup in the linear forms 
$\log\SU(n)^T$ of the Lie algebra
\begin{eqn}{rl}
\ro{grp}~\SU(n)&=\{
\brack{2j_1,\dots,2j_{n-1}}\mid 2j_k\in\Z\}\cr
&=\ro{grp}~\U(1)^{n-1} =\Z^{n-1}\cr
\ro{mon}~\SU(n)&\subnoteq \ro{grp}~\SU(n)
\end{eqn}The integers $\{2j_r\}_{r=1}^{n-1}$ can be related to the 
winding numbers of the
Cartan $\U(1)$'s involved. Since $\SU(n)$ are simple groups, they
come in selfdual $\SO(2)$ re\-pre\-sen\-ta\-tions.
Especially for $\SU(2)$ the halfintegers $(J,j)\in{\N\over2}\x{\Z\over2}$ 
are called spin and
its 3rd component.

The {\it defining} re\-pre\-sen\-ta\-tion 
with a  complex $n$-di\-men\-sio\-nal re\-pre\-sen\-ta\-tion space can 
be written with the generalized Pauli (subsection 2.1)
 matrices 
\begin{eqn}{l}
\brack{1,0,\dots,0}(\al)\cong e^{\al_a {i\si(n)^a\over2}}
\end{eqn}where the Cartan subgroup $\U(1)^{n-1}$ is represented with
the $(n-1)$ diagonal matrices
\begin{eqn}{l}
\U(1)^{n-1}=\{
\exp{\SUM_{k=2}^n}\al_{k^2-1}{i\si(n)^{k^2-1}\over2}\mid\al_{k^2-1}\in\R\}
\end{eqn}Taking together the diagonals of the 
$(n-1)$  matrices $\{{1\over2}\si(n)^{k^2-1}\}_{k=2}^n$, one obtains
the $n$ weights $\{w_r\}_{r=1}^n$
 of the defining $\SU(n)$-re\-pre\-sen\-ta\-tions
in the real $(n-1)$-di\-men\-sio\-nal weight space. 
In the normalization\footnote{\scriptsize
The integer winding number normalization arises with the basis
$\{\sqrt{{k\choose2}}\si(n)^{k^2-1}\}_{k=2}^n$.}
with the Pauli matrices, the 
defining weights  occupy the 
corners of a regular
fundamental simplex, centered at the origin, as expressed by the
$\brack{(n-1)\x n}$-matrix 
\begin{eqn}{l}
\ro{weights}~\brack{1,0,\dots,0}\cong\ro{simplex}~(n)\cr
={\scriptsize\pmatrix{w_1\cr w_2\cr w_3\cr w_4\cr\dots\cr\dots\cr w_n\cr}}
={1\over2}{\scriptsize\pmatrix{
1&{1\over\sqrt3}&{1\over\sqrt6}&\dots&{1\over\sqrt{{n\choose2}}}\cr
-1&{1\over\sqrt3}&{1\over\sqrt6}&\dots&{1\over\sqrt{{n\choose2}}}\cr
0&-{2\over\sqrt3}&{1\over\sqrt6}&\dots&{1\over\sqrt{{n\choose2}}}\cr
0&0&-{3\over\sqrt6}&\dots&{1\over\sqrt{{n\choose2}}}\cr
&\dots&&\dots&\dots\cr
&\dots&&\dots&\dots\cr
0&0&0&\dots&-{n-1\over\sqrt{{n\choose2}}}\cr}} 
,~~\left.\begin{array}{l}
{\SUM_{r=1}^n}w_r=0\cr
\norm{w_k-w_j}=\de_{kj}\cr
\norm{w_k}=\sqrt{{n-1\over 2n}}\end{array}\right. 
\end{eqn}

All complex  {\it finite di\-men\-sio\-nal} irreducible re\-pre\-sen\-ta\-tions of 
the simple Lie group
$\SL(\C_\R^n)$, $n\ge2$,   are characterized 
 - according to the Cartan subgroup $\GL(\C_\R)^{n-1}$ - 
 by $2(n-1)$ positive  integers, interpretable as 
 $(n-1)$ 'left' and $(n-1)$ 'right' winding
 numbers
\begin{eqn}{rl}
\x \ro{mon}_{\ro{fin}}\SL(\C_\R^n)&=
\{\brack{2L_1,\dots,2L_{n-1}|2R_{n-1},\dots,2R_1}\mid
2L_k,2R_k\in\N\}\cr
&=\ro{mon}~\SU(n)\x \ro{mon}~\SU(n)=\N^{n-1}\x\N^{n-1}\cr
\end{eqn}reflecting the  $\N$-linear combinations from  
$2(n-1)$ {\it fundamental} re\-pre\-sen\-ta\-tions
(one $1$, elsewhere $0$). Also the weight group is
the 'square' of the weight group for the unitary subgroup
\begin{eqn}{rl} 
\x \ro{grp}_{\ro{fin}}\SL(\C_\R^n)
&=\{\brack{2l_1,\dots,2l_{n-1}|2r_{n-1},\dots,2r_1}\mid
2l_k,2r_k\in\Z\}\cr
&=\ro{grp}~\SU(n)\x\ro{grp}~\SU(n)=\Z^{n-1}\x\Z^{n-1}\cr
\x \ro{mon}_{\ro{fin}}\SL(\C_\R^n)&\subnoteq\x  \ro{grp}_{\ro{fin}}\SL(\C_\R^n)
\end{eqn}The 
finite dimensional irreducible
$\SL(\C^n_\R)$-representations are not necessarily unitary.
However, the monoid and the group allow 
a {\it conjugation}
\begin{eqn}{rl}
\brack{2L_1,\dots,2L_{n-1}|2R_{n-1},\dots,2R_1}^\x
&=\brack{2R_1,\dots,2R_{n-1}|2L_{n-1},\dots,2L_1}\cr
\brack{2l_1,\dots,2l_{n-1}|2r_{n-1},\dots,2r_1}^\x
&=\brack{2r_1,\dots,2r_{n-1}|2l_{n-1},\dots,2l_1}\cr
\end{eqn}The equivalence classes with repect to this conjugation
characterize the 
equivalence classes of the finite dimensional
irreducible representations of the {\it complex} group $\SL(\C^n)$.

The conjugated pair of the two {\it defining} 
finite di\-men\-sio\-nal $\SL(\C^n_\R)$-re\-pre\-sen\-ta\-tions
 uses the
Pauli matrices, e.g. for $n=2$ 
the left and right handed Weyl re\-pre\-sen\-ta\-tion
$\brack{1|0}$ and $\brack{0|1}$
 \begin{eqn}{l}
\brack{1,0,\dots,0|0,\dots,0,0}
(x,\al)\cong e^{(x_a+i\al_a){\si(n)^a\over2}}\cr
\brack{0,0,\dots,0|0,\dots,0,1}(x,\al)
\cong e^{(- x_a+i\al_a){\si(n)^a\over2}}\cr
\end{eqn}These 
representations are not unitary,
they are equivalent for the complex group $\SL(\C^n)$.

Only the 
self conjugated irreducible 
re\-pre\-sen\-ta\-tions 
are also unitary (indefinite unitary). They define the
representation monoid and weight group for
the adjoint  group $\SL(\C^n_\R)/\I_n$
\begin{eqn}{rl}
\ro{mon}~\SL(\C^n_\R)/\I_n&=
\{\brack{2J_1,\dots,2J_{n-1}|2J_{n-1},\dots,2J_1}\mid 2J_k\in\N\}\cr
&=\N^{n-1}\cr
\ro{grp}~\SL(\C^n_\R)/\I_n&=\Z^{n-1}
\end{eqn}with the $(n-1)$ {\it fundamental} representations
$\brack{1,0,\dots,0| 0,\dots,0,1}$ etc.
The conjugation compatible analogue to the 
real $(n^2-1)$-di\-men\-sio\-nal adjoint re\-pre\-sen\-ta\-tion for
$\SU(n)$
is the irreducible 
unitary re\-pre\-sen\-ta\-tion of $\SL(\C^n_\R)$ 
on a real $n^2$-di\-men\-sio\-nal space, e.g. on the ti\-me\-spa\-ce translations 
$\R(n)\cong\log\GL(\C^n_\R)/\log\U(n)$, faithful for 
$\SL(\C^n_\R)/\I_n$, i.e. for the Lorentz group $\SO^+(1,3)$ 
in the case  $n=2$ 
\begin{eqn}{rl}
\hbox{$\x$-adjoint re\-pre\-sen\-ta\-tion: }
&\brack{1,0,\dots,0|0,\dots,0,1}\cr
\hbox{$n=2$: }&\brack{1|1}\cr
\end{eqn}The  $n^2$-dimensional indefinite unitary irreducible 
$\SL(\C^n_\R)$-representation is called the {\it defining} representation
of the adjoint group $\SL(\C^n_\R)/\I_n$, e.g. the
Minkowski representation of $\SO^+(1,3)$ on $\R(2)$.

Any   re\-pre\-sen\-ta\-tion of $\SL(\C^n_\R)$ is a  re\-pre\-sen\-ta\-tion 
for $\SU(n)$ - in general decomposable -  and gives  - by the quotient of the
represented groups - a  realization of the Sylvester manifold
$\bl{SD}(n)=\SL(\C^n_\R)/\SU(n)$, e.g. for $n=2$
with the Pauli matrices $\rvec\si$ and the Lorentz boost matrices $\rvec B$
\begin{eqn}{l}
\brack{1|0}( x)=e^{{\rvec x\rvec\si\over2}},~~
\brack{0|1}( x)=e^{-{\rvec x\rvec\si\over2}}\cr
\brack{1|1}( x)=e^{\rvec x\rvec B},~~
{\rvec x\rvec B}={\scriptsize\pmatrix{
0& x_1& x_2& x_3\cr
 x_1&0&0&0\cr
 x_2&0&0&0\cr
 x_3&0&0&0\cr}}
\end{eqn}

\subsection{Irreducible Representations of $\SL(\C^n_\R)$}

As shown by Gel'fand and Naimark \cite{NAIM,GELNAI,GELVIL},
all irreducible $\SL(\C^n_\R)$-re\-pre\-sen\-ta\-tions
can be characterized by the irreducible re\-pre\-sen\-ta\-tions of
the Cartan subgroup $\GL(\C_\R)^{n-1}$.

To illustrate the case $n=2$:
The re\-pre\-sen\-ta\-tion spaces of 
\begin{eqn}{l}
\SL(\C_\R^2)=\{\la=
e^{(\rvec x+i\rvec\al){\rvec\si\over2}}
={\scriptsize\pmatrix{\al&\be\cr\ga&\de\cr}}\mid
\det\la=1\}\cr
\end{eqn}are
subspaces of the complex vector space
$\C^{\C^2}=\{f:\C^2\map\C\}$ with the complex valued mappings 
on the vector space $\C^2$. The $\SL(\C^2_\R)$-action is induced by the 
defining re\-pre\-sen\-ta\-tion on $\C^2$
\begin{eqn}{l}
\Diagr{\C^2}{\C^2}\C\C{\la}{_\la f}{\id_\C}f,~~~~~
_\la f (z_1,z_2)
=f (\la^{-1}\m (z_1,z_2))\cr
\la^{-1}\m(z_1,z_2)=(z_1,z_2)\la=
(z_1,z_2){\scriptsize\pmatrix{\al&\be\cr \ga&\de\cr}}=
(\al z_1+\ga z_2, \be z_1+\de z_2)\cr
_{\la'\o\la}f=_{\la'}(_\la f)\cr
\end{eqn}

The irreducible $\SL(\C^2_\R)$-re\-pre\-sen\-ta\-tions
use the irreducible re\-pre\-sen\-ta\-tions
of its Cartan subgroup 
$\GL(\C_\R)$
on the subspace of the mappings which are  
$(\nu_1|\nu_2)$-ho\-mo\-ge\-neous  with repect to  a 'relative'    
$\GL(\C_\R)$ for the two components $z_{1,2}$  
\begin{eqn}{l}
\C^{\C^2}_{(\nu_1|\nu_2)}=\{f\in\C^{\C^2}\mid
f (\de z_1,\de z_2)
=\de^{\nu_1}\ol\de^{\nu_2}f (z_1,z_2)\hbox{ for }\de\in\C\}\cr
\hbox{with }\cases{
\nu_1={i m+Z\over2},~~
\nu_2={i m-Z\over2}\cr
( m;Z)=\(-i(\nu_1+\nu_2);\nu_1-\nu_2\)\in\C\x\Z\cr}\cr
\end{eqn}Those functions involve an integer winding number $\pm Z$
(spin ${Z\over2}$) for the 'relative' phase group $\U(1)\subnoteq \SU(2)$
 and a complex
number $m$ for the 'relative' causal group 
$\D(1)\subnoteq
\bl{SD}(2)=\SL(\C^2_\R)/\SU(2)$.
Since the $(\nu_1|\nu_2)$-ho\-mo\-ge\-neous mappings have the orbit properties
\begin{eqn}{l}
f (z_1,z_2)
=z_2^{\nu_1}\ol{z_2}^{\nu_2} 
f ({z_1\over z_2},1)
\end{eqn}the group $\SL(\C^2_\R)$
acts on the corresponding vector space
\begin{eqn}{l}
F(\ze)=f (\ze,1)\in\C^\C=\{F:\C\map\C\}
\end{eqn}in the following form  
\begin{eqn}{rl}
\la&\mape D^{( m;Z)}(\la),~~F\mape D^{( m;Z)}(\la)(F)= {_\la F}\cr
_\la F(\ze)&=F({\al \ze+\ga\over\be \ze+\de})
|\be \ze+\de|^{i m}({\be\ze+\de\over \ol{\be \ze+\de}})^{{Z\over2}}\cr
&=F({\al \ze+\ga\over\be \ze+\de})
(\be\ze+\de)^{\nu_1}( \ol{\be \ze+\de})^{\nu_2}\cr
\end{eqn}

The pairs $(m;Z)\in\C\x\Z$ characterize all 
irreducible complex re\-pre\-sen\-ta\-tions of $\SL(\C_\R^2)$,
not necessarily unitary.
The {\it finite di\-men\-sio\-nal irreducible} 
$\SL(\C_\R^2)$-re\-pre\-sen\-ta\-tions of the former
subsection
arise with an integer imaginary  'causal number'
$ m$
\begin{eqn}{rl}
(m;Z)\in i\Z\x\Z&\cr 
 \x \ro{grp}_{\ro{fin}}\SL(\C^2_\R)&=\{\brack{2\nu_1|2\nu_2}=
  \brack{2l|2r}\}\cr
  &=\ro{grp}~\SU(2)\x \ro{grp}~\SU(2)=\Z\x\Z
\end{eqn}

As to be expected from the
abelian group $\GL(\C_\R)$, 
the  {\it unitary principal irreducible}
$\SL(\C^2_\R)$-re\-pre\-sen\-ta\-tions are characterized by a
real {\it causal unit $ m$ (mass)}
 \begin{eqn}{rl}
\ro{grp}_{\ro{princ}}\SL(\C_\R^2)&=\{(m;Z)\}\cr
 &=\ro{grp}~\GL(\C_\R)= \R\x \Z\cr
\end{eqn}which reflects the $\U(2)$-unitary representations
(conjugation $\star$)  of the Cartan subgroup 
\begin{eqn}{l}
e^{(x_3+i\al_3){\si^3\over2}}\mape u=e^{(imx_3+iZ\al_3){\si^3\over2}}\cr
(m;Z)\in\R\x\Z\then u^\star=u^{-1} \in\U(1)_3\subnoteq \SU(2)\cr
\end{eqn}

The unitary representations with trivial causal unit $m=0$ are the 
finite dimensional self conjugated representations of the former
subsection
\begin{eqn}{l}
(0;2J)\cong \brack{2J|2J}\end{eqn}Those massless indefinite unitary 
irreducible representations with
the defining 4-dimensional Minkowski representation 
$\brack{2J|2J}=2J\brack{1|1}$ are used for gauge fields in relativistic
field theories.

With repect to the indefinite unitary group $\U(1,1)$
with  conjugation $\x$
\begin{eqn}{l}
{\scriptsize\pmatrix{\al&\be\cr\ga&\de\cr}}^\x=
{\scriptsize\pmatrix{0&1\cr1&0\cr}}
{\scriptsize\pmatrix{\al&\be\cr\ga&\de\cr}}^\star
{\scriptsize\pmatrix{0&1\cr1&0\cr}}
={\scriptsize\pmatrix{\ol\de&\ol\be\cr\ol \ga&\ol\al\cr}}\cr
\end{eqn}diagonal representations of the Cartan subgroup 
in $\SU(1,1)$ have to be
of the form
\begin{eqn}{l}
(m;Z)=(i\rho;0)\in i\R\then
u=e^{-\rho x_3{\si^3\over2}}\in\SU(1,1)
\end{eqn}leading to the  
{\it unitary supplementary irreducible}
$\SL(\C^2_\R)$-re\-pre\-sen\-ta\-tions with trivial winding numbers
and an imaginary causal number $i\rho$
 \begin{eqn}{rl}
\ro{grp}_{\ro{suppl}}\SL(\C_\R^2)&=\{(i\rho;0)\}= i\R\cr
\end{eqn}

The equivalences classes for the
representations in the principal and supplementary series
are given in   \cite{NAIM,GELNAI, GELVIL}. 
For the
{\it positive unitary}
representations\footnote{\scriptsize
Already Gel'fand and Naimark \cite{GELNAI} call the requirement of
positive unitarity for $\SL(\C^n_\R)$-representations
'in a certain sense unnatural'.}
one has to discuss also the scalar products for the
representation spaces, especially for the supplementary series.

The generalizations for $\SL(\C^n_\R)$, $n\ge2$,  with Cartan subgroup
$\GL(\C_\R)^{n-1}$
 are given for the principal representations with
 $(n-1)$ real causal units and $(n-1)$ integer winding numbers 
 \begin{eqn}{rl}
\ro{grp}_{\ro{princ}}\SL(\C_\R^n)&=
\{(m_1,\dots,m_{n-1};Z_1,\dots,Z_{n-1})\}\cr
&=\ro{grp}~\GL(\C_\R)^{n-1} = \R^{n-1}\x \Z^{n-1}\cr
(0,\dots,0;2J_1,\dots,2J_{n-1})&\cong
\brack{2J_1,\dots,2J_{n-1}|2J_{n-1},\dots, 2J_1}\cr
\end{eqn}

The supplementary series has to take into account the diagonal
$\U(1,1)$ structure 
\begin{eqn}{l}
{\scriptsize\pmatrix{
e^{im x+iZ\al}&0\cr
0&e^{-im x+iZ\al }\cr}}\in \U(1,1)\hbox{ for } (m;Z)\in i\R\x\Z
\end{eqn}Therewith the supplementary weights are characterized by
coinciding winding number pairs and conjugated causal numbers
(more details in \cite{NAIM,GELNAI, GELVIL})
\begin{eqn}{l}
(m_1,\dots,m_{n-1};Z_1,\dots,Z_{n-1})\in\C^{n-1}\x\Z^{n_1}\cr
\hbox{ with entries }(m,\ol m;Z,Z)
\end{eqn}

\section{Energy-Momenta Measures}

As to be expected from their Cartan subgroups $\GL(\C_\R)^{n-1}$, $n\ge2$,
also the groups $\SL(\C^n_\R)$ have {\it reducible, but nondecomposable}
representations as first discussed by
Shelobenko \cite{SHELO}. 

Therewith, I suspect\footnote{\scriptsize
A mathematically rigorous classification
of all {\it nondecomposable unitary} representations of $\GL(\C^n_\R)$ for
$n\ge2$ would be appreciated.}  
that the {\it $\GL(\C^n_\R)$-representation monoid with real
causal units}  for
the equivalence classes of all unitary nondecomposable representations
is given by the representation monoid for the Cartan subgroup
\begin{eqn}{rcl}
\ro{mon}_{\ro{princ}}\GL(\C^n_\R)&
\stackrel{?}{=}&
\{(N_1,\dots,N_n|m_1,\dots,m_n;Z_1,\dots,Z_n)\}\cr
&=&\ro{mon}~\GL(\C_\R)^n=\N^n\x\R^n\x\Z^n\cr
\end{eqn}One has for the weight groups
with real causal units \footnote{\scriptsize
With $\U(n)\cong{\U(1)\x\SU(n)\over \I_n}$, one has to take care of the
phase correlations for both unitary factors, e.g. relevant for
the isospin-hypercharge correlation in the standard model \cite{S921}.}
\begin{eqn}{rl}
\ro{grp}_{\ro{princ}}\GL(\C^n_\R)&=
\{(m_1\dots,m_n;Z_1,\dots,Z_n)\}\cr
&=\ro{grp}~\GL(\C_\R)^n=\R^n\x\Z^n\cr
\ro{grp}~\U(n)&=\{\brack{Z_1,\dots,Z_n}\}\cr
&=\ro{grp}~\U(1)^n=\Z^n\cr
\end{eqn}

If such a conjecture is true, one is lead to the suggestion for
the harmonic analysis of the causal 
ti\-me\-spa\-ce manifolds
\begin{eqn}{rcl}
\ro{mon}_{\ro{princ}}\D(n)&\stackrel{?}{=}&
\{(N_1,\dots,N_n|m_1,\dots,m_n)\}\cr
&=&\ro{mon}~\D(1)^n=\N^n\x\R^n\cr
\end{eqn}i.e. 
the nondecomposable realizations of the homogeneous spaces $\D(n)$ 
would be characterized by $n$ natural numbers 
$N_k\in\N$ for the dimensions (discrete invariants)
and $n$ causal units $m_k\in\R$ (continuous invariants).

In this chapter, I shall try  to concretize those structures 
following the analogies to the 
abelian case used for the Cartan subgroups.

\subsection{Algebra of  Causal Measures}

Since  re\-pre\-sen\-ta\-tions  of the causal group are characterized 
by a continuous real invariant
$m$ (mass), it is appropriate to use
a {\it Lebesgue measure} on the Lie algebra linear forms.
The  causal re\-pre\-sen\-ta\-tions $(N|m)$ 
are expressible with Dirac distributions (point measures)
$\de_m\cong\de(m-\mu)$ and their derivatives, e.g.
\begin{eqn}{rl}
   (0|m)(\tau)&= \int d\mu~ \de(m-\mu)e^{i\mu \tau}\cr 
   (1|m)(\tau)&
   = \int d\mu~{\scriptsize\pmatrix{
    \de(m-\mu)&\la\de'(m-\mu)\cr
     0&\de(m-\mu)\cr}}  
   e^{i\mu \tau}\cr 
   (2|m)(\tau)&
   = \int d\mu~{\scriptsize\pmatrix{
    \de(m-\mu)&\la\de'(m-\mu)&{\la^2\over2!}\de''(m-\mu)\cr
    0&\de(m-\mu)&\la\de'(m-\mu)\cr
    0&0&\de(m-\mu)\cr}}  
   e^{i\mu \tau}\cr
   \hbox{etc.}&\cr 
\end{eqn}

A {\it  causal  measure} of the 
mass space $\log\D(1)^T\cong\R$ for the
anlaytic manifold $\D(1)$  
is defined by its property to define a function, analytic in
the causal translations $\tau\in\R$  
\begin{eqn}{l}
e^\tau\mape \int d\mu~h(\mu)e^{i\mu\tau}\cr
\end{eqn}

A measure can be multiplied with a complex number.
Two measures  can be added  and multiplied via   
the $\de$-additive convolution,
induced by the composition in the re\-pre\-sen\-ta\-tion monoid 
\begin{eqn}{l}
(h*h')(\mu)
=\int d\mu_1 d\mu_2 
h(\mu_1)\de(\mu_1+\mu_2-\mu)h'(\mu_2)\cr
\end{eqn}Therewith
the causal  measures $\meas \R$ 
have the structure of 
an {\it abelian unital algebra} with the
 unit given by the underived Dirac measure
for trivial frequency (mass) $\de_0$. 

 A causal measure 
 $h_{N}$ has  the {\it momentum} $ N\in\N$ if it obeys
  the conditions
\begin{eqn}{rl}
 N=0:&\meas_0 \R=\{ h_0\mid \int d\mu ~ h_0(\mu)\ne 0\}\cr
 N\ge1:&\meas_{N} \R=\{ h_{N}\mid
\cases{
\int d\mu ~\mu^k h_{N}(\mu)=0,~~k=0,\dots,N-1\cr
\int d\mu ~\mu^{N} h_{N}(\mu)\ne0\cr}   \}\cr
\end{eqn}with the Dirac point measures as examples
\begin{eqn}{l}
\de^{(N)}(m-\mu)\cong\de^{(N)}_m\in\meas_{N}\R
\end{eqn}The convolution multiplication is 
compatible with the momentum properties
\begin{eqn}{l}
\meas_{N_1}\R *\meas_{N_2}\R\sub\meas_{N_1+N_2}\R\cr
\end{eqn}

The normalization $\int d\mu~h(\mu)$ of a causal measure   
reflects the  re\-pre\-sen\-ta\-tion of the causal group unit $1\in\D(1)$.
Its first causal momentum will be called  
\begin{eqn}{l}
\hbox{\it causal unit : }m=\int d\mu~\mu h(\mu)
\end{eqn}

With respect to the possibly indefinite unitary causal re\-pre\-sen\-ta\-tions,
e.g. $(1|m)$ in $\U(1,1)$, the measures are not required to be 
positive definite.
This feature has to be taken care of in the 
probability interpretation of quantum theories. It 
 will be discussed in connection with the spacelike supported
parts of a propagator, i.e. with respect to the in- and outgoing
particle interpretable causal re\-pre\-sen\-ta\-tions (subsection 5.3).

The functions arising in the quantization of linear fields (subsection 1.2)
have the $\D(1)$-analysis 
\begin{eqn}{l} 
\cl E_n(\tau^2)=
\int d\mu~h_0^{(n)}(\mu)e^{i\mu\tau}\cr
h_0^{(n)}(\mu)={1\over B({1\over2},{1\over2}+n)}~
\vth(1-\mu^2)(1-\mu^2)^{n-{1\over2}},~~\int d\mu~h_0^{(n)}(\mu)=1\cr
\end{eqn}

Any dynamics  determines a subalgebra of the causal algebra, e.g.:
The  subalgebra $\meas_0\R$ 
is related to the measures for irreducible causal re\-pre\-sen\-ta\-tions.
Its subalgebra $\log\D(1)^T\cong \R$ uses only  the point supported 
Dirac measures $\log\D(1)^T\cong \{\de_m\mid m\in\R\}$ - it 
is the algebra for  the irreducible  $\U(1)$-re\-pre\-sen\-tations
of the causal group (positive unitary characters). Its 
discrete subgroup with
the integers $\{\de_{zm}\mid z\in\Z\}\cong\Z$ is used
for the time development of a harmonic oscillator with frequency
$m$. Any subset of the causal measure algebra together with the unit 
generates a subalgebra.  Such a 
set may be used in a perturbative approach to generate 
the full subalgebra, associated to the dynamics to be solved.
E.g. the  
2-elementic set $\{\de_{\pm m}|\mid m\ne0\}$ , associated to the
irreducible re\-pre\-sen\-ta\-tions $e^{\pm im\tau}$,  
generates the 
causal subgroup $\Z$ and the associated re\-pre\-sen\-ta\-tions 
for the quantum oscillator.

It is straightforward to define the algebra $\meas \R(n)^T$ of causal measures
\begin{eqn}{l}
\int d^{n^2}q ~h(q) e^{iqx}
\end{eqn}for general $n\ge1$ and energy-momenta $q\in\R(n)^T$.
The Lebesque measures $d^{n^2}q$ are invariant under 
the affine group $\SL(\R^{n^2})\x_s\R^{n^2}$
and, therewith, Poincar\'e invariant.

\subsection{Representation Structure of Linear Fields}

Linear field theories on 
the 4-di\-men\-sio\-nal ti\-me\-spa\-ce $\R(2)$ (subsection 1.2)
are not compatible with the  algebra structure of causal measures.
Fields with distributive quantization cannot be used as a generating set. 
The divergencies in Feynman integrals, e.g. in the vacuum polarization of
quantum electrodynamics, involving the undefined product
$\brack{\ga_k\bl c^k(m|x)+i\bl s(m|x)}^2$,
show that the convolution  product does not make sense
for linear fields.

In the distributive quantization
of quantum fields (subsection 1.2), one  uses 
the naive analogue of the 
1-di\-men\-sio\-nal time $\D(1)$ structure 
\begin{eqn}{rl}
(0||m^2)(t)&
\cong {\scriptsize\pmatrix{
\cos mt &{i\over m}\sin mt  \cr im \sin mt  &\cos mt \cr}}\cr
{\scriptsize\pmatrix{
\cos m t\cr 
i\sin mt\cr}}
&=\int dq_0
 \de(m^2-q_0^2)\ep(q_0){\scriptsize\pmatrix{q_0\cr m\cr}}
 e^{iq_0t}\cr
\end{eqn}given by the Dirac measure of the energy-momenta
and a frequency $q_0$ analysis 
\begin{eqn}{l}
d^4 q~\de(m^2-q^2)\ep(q_0)=d^3q~dq_0{\de(q_0-\sqrt{m^2+\rvec q^2})
+\de(q_0+\sqrt{m^2+\rvec q^2})\over 2q_0}
\end{eqn}The 
measure, integrated  with an
irreducible  re\-pre\-sen\-ta\-tion $e^{iqx}\in\U(1)$
of the additive translation group $x\in\R(2)$,
describes the quantization of linear particle fields 
\begin{eqn}{rl}
((0||m^2))(x)&=
{\scriptsize\pmatrix{
\bl c^k(m|x)&i\bl s(m|x)\cr i\bl s(m|x)&\bl c^k(m|x)\cr}}\cr
{\scriptsize\pmatrix{
\bl c^k(m|x)\cr
i\bl s(m|x)\cr}}
&=\int {d^4 q\over(2\pi)^3}~
\de(m^2-q^2)~\ep(q_0){\scriptsize\pmatrix{
q^k\cr m \cr}}
e^{iqx}\cr
\end{eqn}These integrated translation re\-pre\-sen\-ta\-tions 
 are space distributions of 
re\-pre\-sen\-ta\-tions of the causal group $\D(1)$
\begin{eqn}{l}
\D(1)\ni e^{x_0}\mape\int d^3x((0||m^2))(x)=   (0||m^2)(x_0)=
{\scriptsize\pmatrix{
\de^k_0\cos mx_0&i\sin mx_0 \cr i\sin mx_0 &\de^k_0\cos mx_0\cr}}\cr
\end{eqn}

Such a time oriented  analysis and - in the Poin\-ca\-r\'e group -
the Wigner classification
is appropriate for  the particle interpretation of a
field theory (subsection 5.3). 
With respect to the ti\-me\-spa\-ce manifold $\D(2)=\D(1)\x\bl{SD}(2)$,
the re\-pre\-sen\-ta\-tion of the Sylvester factor 
$\bl{SD}(2)\cong \SO^+(1,3)/\SO(3)$ is not adequately taken into account  - 
this negligence is the main reason for
the divergency problems working with linear quantum fields.

The realization of the  Cartan group $\D(1)\cong\SO^+(1,1)$ in the 
Sylvester factor   for linear fields can be seen
for  the manifolds $\D(1)\x\SO^+(1,s)/\SO(s)$ as follows: 
The $\SO^+(1,s)$ scalar contribution in the quantization of linear fields
(subsection 1.2) 
\begin{eqn}{l}
i\bl s(m|x)
=\int{d^{1+s}q\over (2\pi)^s}~
\de(m^2-q^2)\ep(q_0) m e^{iqx} 
= {\ep(x_0)\over \pi i}\int{d^{1+s}q \over (2\pi)^s}
m{e^{iqx}\over (q^2- m^2)_P}
\end{eqn}shows the time $\D(1)$ re\-pre\-sen\-ta\-tion properties by the
space integral 
\begin{eqn}{l}
\int d^s x~\bl s(m|x)=\sin m x_0
\end{eqn}whereas the realization of $\D(1)$ in the Sylvester factor for $s\ge1$ 
shows up in the ordered time integral
\begin{eqn}{rl}
\int dx_0\ep(x_0)\ep(m)\bl s(m|x)
&=2|m|\int {d^sq\over(2\pi)^s}
{e^{-i\rvec q\rvec x}\over \rvec q^2+m^2}
=\cases{
e^{-|m\rvec x|} &for $s=1$\cr
{|m|\over2\pi |\rvec x|}e^{-|m\rvec x|} &for $s=3$\cr}
\end{eqn}For $s=1$,
the orthogonal manifold 
$\D(1)\x\SO^+(1,s)/\SO(s)$ is isomorphic to the group $\D(1)\x\D(1)$
with elements $e^{x_0-|\rvec x|}$. Here
the exponential $e^{-|m\rvec x|}$ represents the space 
$\D(1)\ni e^{-|\rvec x|}$.
For $s=3$, the Yukawa potential is no re\-pre\-sen\-ta\-tion of 
$\D(1)\subnoteq\bl{SD}(2)\cong\SO^+(1,3)/\SO(3)$.

 \subsection{Point Measures for Energy-Momenta}

For the generalization of the  nondecomposable  re\-pre\-sen\-ta\-tions
$(N|m)$ of the abelian group $\D(1)$ 
to  realizations $(N_1,\dots,N_n|m_1,\dots,m_n)$ of 
the ho\-mo\-ge\-neous ti\-me\-spa\-ces  $\D(n)$ with the Cartan subgroups $\D(1)^n$,
 it is convenient to use the residues
of  loop integrals for the linear forms $\R(n)^T$ (energy-momenta)
of the ti\-me\-spa\-ce translations $\R(n)$
\begin{eqn}{l}
{(i\la\tau)^k\over k!}e^{im\tau}
=\int d\mu~{\la^k\over k!}\de^{(k)}(m-\mu)e^{i\mu\tau}
={1\over2\pi i}\oint d\mu~{\la^k\over(\mu-m)^{1+k}}e^{i\mu\tau},~~
\end{eqn}with a nontrivial constant $\la\in\R$.
Therewith one has as integrands for
the $(1+N)$ elements of a $\D(1)$-re\-pre\-sen\-ta\-tion
\begin{eqn}{rl}
\D(1)\ni e^\tau\mape &(N|m)(\tau)\cr
& d\mu~{\la^k\over(\mu-m)^{1+k}}~e^{i\mu\tau}
\hbox{ with }k=0,\dots,N\cr
\end{eqn}E.g.,  the fundamental complex 2-di\-men\-sio\-nal
 $\D(1)$-re\-pre\-sen\-ta\-tions use both poles and dipoles
\begin{eqn}{l}
(1|m)(\tau)\cong{\scriptsize\pmatrix{
1&i\la\tau\cr
0&1\cr}}
e^{i m\tau}
={1\over2\pi i}\oint{ d\mu\over\mu-m}
{\scriptsize\pmatrix{
1&{\la\over\mu-m}\cr
0&1\cr}}
e^{i\mu\tau}
\end{eqn}

The structure of the poles, i.e. the location and the order 
of the singularities, reflect the continuous and the discrete invariant of the
$\D(1)$-re\-pre\-sen\-ta\-tion $(N|m)$. The irreducible 
re\-pre\-sen\-ta\-tions have the 
{\it irreducible measures} ${d\mu\over \mu-m}$, an additional 
nontrivial $\tau$-dependence is expressed
by the {\it nondecomposable measures} 
${d\mu~\la^k\over (\mu-m)^{1+k}}$, reducible for  $k=1,\dots,N$.

The re\-pre\-sen\-ta\-tion elements of $\D(1)^n$ with Cartan coordinates
$\{\xi_r\}_{r=1}^n$ are  products of loop integrals
\begin{eqn}{rl}
\D(1)^n\ni e^\xi= 
{\scriptsize\pmatrix{
e^{\xi_1}&\dots&0\cr
&\dots&\cr
0&\dots&e^{\xi_n}\cr}}
\mape& 
(N_1|m_1)(\xi_1)\ox \cdots\ox (N_n|m_n)(\xi_n)\cr
{(i\la\xi_1)^{k_1}\cdots (i\la\xi_n)^{k_n}
\over k_1!\cdots k_n!}e^{i(m_1\xi_1+\dots+m_n\xi_n)}
&={1\over(2\pi i)^n}\oint d^n\mu~
{\la^{k_1+\cdots+k_n}e^{i(\mu_1\xi_1+\dots+\mu_n\xi_n)}\over
(\mu_1-m_1)^{1+k_1}
\cdots (\mu_n-m_n)^{1+k_n}}\cr
&\hbox{ with }\cases{
k_r=0,\dots,N_r\cr r=1,\dots,n\cr}\cr
\end{eqn}

As illustrated in the former subsection,
the conventional relativistic field quantization for $n=2$  uses only 
$\D(1)$-re\-pre\-sen\-ta\-tions
of the determinant in $\U(1)$
\begin{eqn}{l}
\D(1)\ni \det e^\xi=e^{\xi_1+\cdots+\xi_n}=e^{x_0}\mape e^{im x_0}\in\U(1)
\end{eqn}The correspondingly frugal analogue for the 
compact group $\U(n)$ are its $\U(1)$-re\-pre\-sen\-ta\-tions 
with only one winding number $Z\in\Z$  
\begin{eqn}{l}
\U(n)\sup\U(1)^n\ni e^{i\be}=
{\scriptsize\pmatrix{
e^{i\be_1}&\dots&0\cr
&\dots&\cr
0&\dots&e^{i\be_n}\cr}}\mape
e^{iZ(\be_1+\cdots+\be_n)}\in\U(1) 
\end{eqn}unfaithful for $n\ge2$.

If the Cartan subgroup 
$\D(1)^n$ is realized in the ho\-mo\-ge\-neous ti\-me\-spa\-ce $\D(n)$
\begin{eqn}{l}
\D(1)\to \D(1)^n\inmap \D(n)
\end{eqn}using Lebesque
measures for the tangent structures
\begin{eqn}{l}
d\mu\hbox{ on }\R\to d^n\mu=d\mu_1\cdots d\mu_n\hbox{ on }\R^n
\inmap d^{n^2}q \hbox{ on }\R(n)^T
\end{eqn}the $n$ Cartan masses $(m_1,\dots,m_n)$ 
for the $\GL(\C^n_\R)$-weights come as poles of
the $\SL(\C^n_\R)$-invariant determinant $\det q=q^n$
with the $n$th powers $m^n$ of the causal masses
\begin{eqn}{l}
{d\mu\over \mu-m}e^{i\mu\tau}\to 
{d^n\mu  \over (\mu_1-m_1)\cdots(\mu_n-m_n)}e^{i\mu_i\xi_i}\inmap 
{d^{n^2} q \over (q^n-m_1^n)\cdots (q^n-m_n^n)}e^{iqx}
\end{eqn}

In the conventional field quantization for $n=2$  only one continuous invariant 
is used (subsection 1.2) 
\begin{eqn}{l}
{d\mu\over \mu-m}e^{i\mu x_0}\inmap
{d^{n^2} q \over q^n-m^n}e^{iqx}
\end{eqn}

The invariance group of the 
{\it irreducible point measures} for the $\D(n)$-rea\-li\-za\-tions
\begin{eqn}{l}
{d^{n^2} q \over (q^n-m_1^n)\cdots (q^n-m_n^n)},~~(m_1,\dots,m_n)\in\R^n
\end{eqn}is  the adjoint group $\SL(\C^n)/\I_n$.
The singularities arise for the invariant energy momenta
\begin{eqn}{l}
\mu(q)=\ep(q)\Big||\det q|^{{1\over n}}\Big|=\cases{
\mu\hbox{ for }n=1\cr
\ep(q_0)\vth(q^2)\sqrt{q^2}\hbox{ for }n=2\cr}
\end{eqn}i.e. one real pole for  odd rank $n$ of ti\-me\-spa\-ce and two real poles
with opposite sign for even rank
\begin{eqn}{l}
\mu^n-m^n=(\mu-m)(\mu^{n-1}+m\mu^{n-2}+\cdots+m^{n-1})\cr
\{\mu(q)\in\R\mid \mu(q)^n=m^n=\left\{\begin{array}{ll}
\{m\}&\hbox{for }n=1,3,\dots\cr
\{\pm m\}&\hbox{for }n=2,4,\dots\end{array}\right\}\sub m\I_n
\end{eqn}Therewith for  even rank, e.g. for the relativistic case, 
the re\-pre\-sen\-ta\-tions
are quadratic $m^2$-dependent.

The nonabelian compact properties
in $\D(n)\cong\D(1)^n\x\SU(n)/\U(1)^{n-1}$, nontrivial for $n\ge2$,
have to be realized via tensor product  polynomials 
$q\ox\cdots \ox q$ of degree $2J\in\N$
in the energy-momenta, placed  in the numerator of the integrand
\begin{eqn}{rl}
{\la^k\over (\mu-m)^k}
\to 
{\la^{k_1+\cdots+k_n}\over
(\mu_1-m_1)^{1+k_1}\cdots (\mu_n-m_n)^{1+k_n}}
\then &{\la^{k_1+\cdots+k_n}~(q\ox\cdots\ox q)_{2J~\ro{times}}
\over (q^n-m^n_1)^{1+k_1}\cdots (q^n-m^n_n)^{1+k_n}}\cr
\hbox{with}& 2J=(n-1)(k_1+\dots+ k_n)\cr
\end{eqn}

Therewith the relevant integrands for the realization of 
a ti\-me\-spa\-ce point 
$d\in \D(n)$ are given with its translations 
$x=\log d\in\R(n)$ 
\begin{eqn}{l}
\D(n)\ni e^x\mape (N_1,\dots,N_n|m_1,\dots,m_n)(x)\cr
d^{n^2}q
{\la^{k_1+\cdots+k_n}(q\ox\cdots\ox q)_{2J~\ro{times}}
\over (q^n-m^n_1)^{1+k_1}\cdots (q^n-m^n_n)^{1+k_n}}e^{iqx}
\hbox{ with }\cases{
k_r=0,\dots,N_r\cr 
r=1,\dots,n\cr
2J=(n-1)(k_1+\cdots+k_n)\cr
}\cr
\end{eqn}The discrete invariants can be taken in the order
\begin{eqn}{l}
N_1\le N_2\le\dots\le N_n
\end{eqn}

\subsection{Realizations of Relativistic Timespace}

The re\-pre\-sen\-ta\-tions of the rank 2  
relativistic ti\-me\-spa\-ce manifold $\D(2)\cong\D(1)^2\x\SU(2)/\U(1)$ 
involve two Cartan masses $(m_1, m_2)\in\R^2$ for the Cartan subgroup $\D(1)^2$
\begin{eqn}{l}
\D(2)\ni e^x\mape (N_1,N_2| m_1,m_2)(x)\cr
d^4q~{\la^{k_1+k_2} (q\ox\cdots\ox q)_{(k_1+k_2)~\ro{times}}\over 
(q^2-m^2_1)^{1+k_1}(q^2-m^2_2)^{1+k_2}}~e^{iqx}
\hbox{ with }\cases{
k_1=0,\dots,N_1\cr 
k_2=0,\dots,N_2\cr 
}\cr
\end{eqn}

The scalar  re\-pre\-sen\-ta\-tions use the 
irreducible scalar measures
\begin{eqn}{ll}
\D(2)\ni e^x\mape &(0,0| m_1,m_2)(x)\cr&
 {d^4q\over 
(q^2-m^2_1)(q^2-m^2_2)}~e^{iqx}\cr
\end{eqn}leading to the well behaved explicit functions (subsection 1.2)
\begin{eqn}{ll}
(0,0|m_1,m_2)(x)
&={\ep(x_0)\over \pi }\int {d^4q\over\pi^2}
{1 \over (q^2-m^2_1)_P(q^2-m^2_2)_P}e^{iqx}\cr
&=\ep(x_0)\vth(x^2)
{m_1^2\cl E_1(m_1^2x^2)-m_2^2
\cl E_1(m_2^2 x^2)\over m_1^2-m_2^2}\cr
(0,0|m,m)(x)&=\ep(x_0)\vth(x^2)\cl E_0(m^2x^2)\cr
(0,0|0,0)(x)&=\ep(x_0)\vth(x^2)\cr
\end{eqn}The principal value pole integration (denoted by $P$)  
has been used. The massless case with the trivial realization  
is particularly simple.

The reducible, but nondecomposable realizations with $(N_1,N_2)=(0,1)$
are faithful and nontrivial for
both the abe\-lian group $\D(1)$ and
the nonabelian boost manifold $\SL(\C^2_\R)/\SU(2)$  - they 
involve both poles and dipoles
\begin{eqn}{rl}
\D(2)\ni e^x\mape& (0,1|m_1,m_2)(x)\cr&
{d^4q\over (q^2-m^2_1)(q^2-m^2_2)}{\scriptsize\pmatrix{
1&{\la q\over q^2-m^2_2}\cr
0&1\cr}}e^{iqx}
\end{eqn}Here one has  the explicit functions
with the reducible measure
\begin{eqn}{l}
{\ep(x_0)\over \pi }
\int {d^4q\over\pi^2}
{\la q \over (q^2-m^2_1)_P(q^2-m^2_2)^2_P}e^{iqx}\cr
~~~=\ep(x_0)\vth(x^2){ix\la \over4}
\Brack{
{m_1^4\cl E_2(m_1^2x^2)-m_2^4
\cl E_2(m_2^2 x^2)\over (m_1^2-m_2^2)^2}-
2{m_2^2\cl E_1(m_2^2x^2)\over  m_1^2-m_2^2}   }\cr
\end{eqn}and the special cases for coinciding and trivial masses
\begin{eqn}{ll}
(0,1|m,m)(x)&=\ep(x_0)\vth(x^2){\scriptsize\pmatrix{
1&{i\la x\over4}\cr 0&1\cr}}\cl E_0(m^2x^2)\cr
(0,1|0,0)(x)&=\ep(x_0)\vth(x^2){\scriptsize\pmatrix{
1&{i\la x\over4}\cr 0&1\cr}}\cr
\end{eqn}

\section{Fundamental Quantum Fields}

In quantum structures,  re\-pre\-sen\-ta\-tions of causal ti\-me\-spa\-ce manifolds
$\D(n)$ are parametrized by operators, i.e. quantum variables:
time dependent positions and momenta 
in quantum mechanics, ti\-me\-spa\-ce dependent
fields in relativistic quantum field theory.

Nondecomposable representations 
$(N|m)$ of the causal group $\D(1)$ are parametrizable
by principal vectors, the irreducible ones $(0|m)$ in $\U(1)$ even by 
eigenvectors (subsection 1.1) 
There exist pairs of {\it cyclic principal vectors} 
$(\ro b,\ro b^\x)$ in the representation space and its dual $V,V\cong\C^{1+N}$, 
which are $\U(N_+,N_-)$-conjugated to each other and parametrize the 
characteristic matrix element\footnote{\scriptsize
With $v\in V$ and $\om\in V^T$ the shorthand notation 
$\dprod{\om(\tau_2)}{v(\tau_1)}=\dprod\om v(\tau_1-\tau_2)$
for the time dependent
dual product is used.}
of the representation
\begin{eqn}{rl}
\dprod{\ro b^\x}{\ro b}(\tau)&={(i\la\tau)^N\over N!} e^{im\tau}\cr
\hbox{e.g. }N=0:&
\dprod{\ro b^\x}{\ro b}(\tau)
=\dprod{\ro u^\star}{\ro u}(\tau)
=e^{im\tau}\cr
N=1:&
\dprod{\ro b^\x}{\ro b}(\tau)
=i\la\tau e^{im\tau}\cr
\end{eqn}Only for the irreducible 
representations, the cyclic vectors are unique;
in general, they are 'gauge dependent' \cite{S911,S923} .

Generalized for ti\-me\-spa\-ces $\D(n)$ with $n\ge1$,
such pairs of cyclic principal vectors  - used as
operators \cite{S922}
 - will be called {\it fundamental ti\-me\-spa\-ce quantum operators}.

\subsection{The Fundamental Mechanical Pair}

Quantum mechanics, $n=1$, uses 
one kind\footnote{\scriptsize
The basically unimportant 
number of fundamental pairs
$\com{iP_a}{X_b}=\de_{ab}, ~a,b=1,\dots,N$, e.g. $N=3$ for
the 3-di\-men\-sio\-nal isotropic oscillator or the 
quantum mechanical nonrelativistic hydrogen atom,
give rise to decomposable $\D(1)$ 
re\-pre\-sen\-ta\-tions with 'internal' degrees of freeedom, e.g. with respect 
to  
$\SU(3)$ (oscillator) and $\SO(4)$ (bound states for the atoms)
resp.}
of  fundamental  pairs - 
a creation-annihilation pair  $(\ro u,\ro u^\star)$ in the complex 
or  a position-momentum pair $(X,iP)$ in the 'real' formulation - 
for the causal group $\D(1)$-re\-pre\-sen\-ta\-tions
in $\U(1)$ and  $\SO(2)$ resp. 
\begin{eqn}{l}
\D(1)\ni e^ t \mape \left\{
\begin{array}{rcl}
\com{\ro u^\star}{\ro u}(t)= \!\!\!&e^{imt}\!\!\!&=(0|m)( t)\cr
{\scriptsize\pmatrix{
\com{iP}X& \com{X}X\cr
\com{P}P&\com{X}{-iP}\cr}}(t)
=\!\!\!&{\scriptsize\pmatrix{
\cos m t&{i\over M m}\sin m t\cr 
iM m \sin m t&\cos m t\cr}}\!\!\!&\cong (0|| m^2)( t)\end{array}\right.
\end{eqn}The   $\D(1)$ re\-pre\-sen\-ta\-tion is generated by the 
Hamiltonian $H={P^2\over2M}+ \ka{X^2\over 2}$.

Re\-pre\-sen\-ta\-tions of $\D(1)$ with a normalized 
 positive frequency measure 
\begin{eqn}{rl}
\D(1)\ni e^ t &  \mape \left\{\begin{array}{rl}
\angle{\com{\ro u^\star}{\ro u}}(t)\!\!\!&=\int d\mu~h_0(\mu)~ e^{i\mu t}\cr
{\scriptsize\pmatrix{
\angle{\com{iP}X}& \angle{\com{X}X}\cr
\angle{\com{P}P}&\angle{\com{X}{-iP}}\cr}}(t)
\!\!\!&=\int d\mu^2~ h_0(\mu^2){\scriptsize\pmatrix{
\cos\mu t&{i\over M\mu}\sin\mu t\cr 
iM\mu \sin\mu t&\cos\mu t\cr}}\end{array}\right.\cr
&\cr
\hbox{for } t=0&:\left\{\begin{array}{rlll}
\angle{\com{\ro u^\star}{\ro u}}(0)&=
\com{\ro u^\star}{\ro u}&=
\int d\mu~h_0(\mu)&=1 \cr
\angle{\com{iP}X}(0)&=\com{iP}X&=
\int d\mu^2~h_0(\mu^2)&=1\end{array}\right.
\end{eqn}arise for the time developments of the
ground state values of the commutators in the case of Hamiltonians 
$H={P^2\over2M}+V(X)$ which lead to bound states only.

There may also occur indefinite unitary  nondecomposable
 re\-pre\-sen\-ta\-tions of the causal group,
if there arise not only bound states, e.g. for a free mass point with
Hamiltonian $H={P^2\over 2M}$
\begin{eqn}{l}
\D(1)\ni e^ t \mape 
{\scriptsize\pmatrix{
\com{iP}X& \com{X}X\cr
\com{P}P&\com{X}{-iP}\cr}}(t)
\cong{\scriptsize\pmatrix{
1&{it\over M}\cr 
0&1\cr}}
\cong (1 |0)( t)\in\U(1,1)\cr
\end{eqn}

\subsection{The Fundamental Quantum Fields}

Many linear fields are used for the  Min\-kow\-ski  
translations $\R(2)$, e.g.
lepton, quark and gauge fields in the standard model of elementary particles.
Those fields, appropriate for a free theory,  do not parametrize 
realizations 
$(N_1,N_2|m_1,m_2)$ of the causal ti\-me\-spa\-ce manifold  $\D(2)$, but
space distributions of time $\D(1)$ re\-pre\-sen\-ta\-tions (subsection 4.2) 
with possible spin degrees of freedom  \cite{S96}. 

E.g., a massive  Dirac field 
(subsection 1.2) 
parametrizes the space distribution of 
the $\U(1)$ time  re\-pre\-sen\-ta\-tion 
\begin{eqn}{rl}
\D(1)\ni e^{x_0}\mape\acom{\ol{\bl\Psi}}{\bl\Psi}(x)
&=\bl{exp}(im|x)=\ga_k\bl c^k(m|x)+i\bl s(m|x)\cr
\hbox{for }x_0=0:~~
\acom{\ol{\bl\Psi}}{\bl\Psi}(\rvec x)&=\ga_0\de(\rvec x)\cr
\end{eqn}

In analogy to the fundamental bosonic pairs $(\ro u,\ro u^\star)$
(quantum mechanics)
with one causal unit $m$,  
{\it fundamental  fermionic pairs}\footnote{\scriptsize
For a full fledged  parametrization,
also $\U(2)$ internal 
hyperisospin degrees of freedom have to be introduced 
to take care of possibly nontrivial properties of the coset $\U(2)$ 
in the ti\-me\-spa\-ce manifold $\GL(\C^2_\R)/\U(2)$.} 
$(\ro b^A,\ro b_{\dot A}^\x)_{A=1,2}$ 
with fundamental 
conjugated $\SL(\C^2_\R)$-representations $\brack{1|0}$ and $\brack{0|1}$ 
are proposed \cite{HEI} for the ti\-me\-spa\-ce manifold
$\D(2)$ (quantum field theory) with two causal units
$m_{1,2}$. They are 
assumed to have as dual product 
(quantization) a faithful $\D(2)\cong \D(1)\x\SL(\C^2_\R)/\SU(2)$ 
realization $(0,1|m_1,m_2)$
\begin{eqn}{l}
\D(2)\ni e^x\mape (0,1|m_1,m_2)(x)\cr
\hbox{with }
\acom{ \ro b^\x}{\ro  b}(x)=
{\ep(x_0)\over \pi i}
\int {d^4q\over\pi^2} {q \over(q^2-m_1^2)_P(q^2-m_2^2)_P^2} e^{iqx}
\end{eqn}

{\bf The  characteristic property of the fundamental fields
is the re\-pre\-sen\-ta\-tion of the external and internal 
operations, used for the definition of the
ti\-me\-spa\-ce manifold $\D(2)=\GL(\C^2_\R)/\U(2)$, not
some linear field equation.} The tangent flat ti\-me\-spa\-ce  interpretation with
linear  particle fields, i.e. with positive unitary re\-pre\-sen\-ta\-tions
of the  Poin\-ca\-r\'e group \cite{WIG}
$\SO^+(1,3)\x_s\R(2)$ for a given dynamics
and the determination of a measure $h(\mu_1^2,\mu_2^2)$
for  a decomposable 
realization - in analogy to $h(\mu^2)$ in the former subsection -
 is the essential part of the solution of a dynamics.
 
A parametrization  of the causal measure with point supported 
realizations
may be useful for a perturbative approach to generate
the causal subalgebra associated with a given dynamics.
Being part of the causal measure algebra, the product re\-pre\-sen\-ta\-tions 
are well defined, 
i.e. there arise no divergencies.

\subsection{Particles in  Fundamental  Fields}

As familiar from regularization recipes,
the price to be paid for the relativistic causal  re\-pre\-sen\-ta\-tions 
with convolution properties (no divergencies) is
the indefinite metric for some re\-pre\-sen\-ta\-tions of the ti\-me\-spa\-ce
translations $\R(2)$,
i.e. an indefinite conjugation, e.g. $\U(1,1)$ (subsection 3.1).  
This price would be too high if it invalidated
 a probability interpretation
of the experimental consequences.
A closer analysis of the manifold $\D(2)$
and its tangent Minkowski translations $\R(2)$ 
reveals a  rather subtle situation:   
The causal und the spacelike
submanifold $\R(2)_\caus$ 
and $\R(2)_\Space$ resp. (subsection 2.1) are reflected in  the 
two parts of Feynman propagators
(subsection 1.2).  
The field quantization realizes the ti\-me\-spa\-ce
manifold $\D(2)$  and is supported
by the  causal submanifold of flat ti\-me\-spa\-ce $\R(2)$.
The  asymptotic consequences
of a dynamics, however,  are interpreted  with the  spacelike submanifold
for in- and outgoing particles. The spacelike submanifold arises 
only as part of the tangent translations $\R(2)$
on the ti\-me\-spa\-ce manifold $\D(2)$ (subsection 2.1).

The fundamental quantum fields introduced in subsection 5.2 
cannot be expanded in terms of energy-momenta
eigenvectors only,
i.e.  with positive $\U(1)$-re\-pre\-sen\-ta\-tions  
of the ti\-me\-spa\-ce translations $\R(2)$.
The causal spreading
on four dimensions leads to negative and derived Dirac measures 
of 
the energy-momenta 
$\R(2)^T$, e.g. $\de'(m^2_2-q^2)$, 
related to reducible, but nondecomposable representations
used for the 'hyperbolic stretching' group $\SO^+(1,1)\cong\D(1)$. 
Therefore, the  fundamental fields
contain, on the one side, positive definite 
$\U(1)$-re\-pre\-sen\-ta\-tions of the 
ti\-me\-spa\-ce translations, describing particles,
e.g. leptons, on the other side they involve 
indefinite, e.g. $\U(1,1)$-re\-pre\-sen\-ta\-tions of the translations
which describe interactions via fields  without particle content.
Examples for nonparticle fields are the Coulomb interactions
(no energy eigenvectors) in the 
four component electromagnetic field \cite{NAK} 
$\bl A^k(x)$ which has
only two particle degrees of freedom (the photons
with left and right handed polarization).
The quantization 
of the electromagnetic field involves 
$\U(1)$ and $\U(1,1)$ re\-pre\-sen\-ta\-tions
of the translations  \cite{SBH95} 
as reflected by the underived and derived point measures
\begin{eqn}{l}
\com{\bl A^k}{\bl A^j}(x)=
\int{d^4q\over (2\pi)^3}
\brack{-e_0^2\eta^{jk}\de(q^2)-\la_0q^kq^j\de'(q^2)}\ep(q_0)e^{iqx}
\end{eqn}with ${e_0^2\over4\pi}$ the fine structure constant and  $\la_0$
a gauge fixing parameter. The Fadeev-Popov ghosts are another example for 
indefinite unitary re\-pre\-sen\-ta\-tions of the translations \cite{S96}.

The spacelike 
(asymptotic) behaviour of a particle field 
$\bl\Psi$ (subsection 1.2) is completely given by
the Fock value of the 'quantization-opposite' commutator,
e.g. for the electron  field
\begin{eqn}{rl}\hbox{for }x^2<0:~~
\angle{\cl C\ol{\bl\Psi}\bl\Psi }(x)&=
\angle { \com{\ol{\bl\Psi}}{\bl\Psi }}(x)\cr
&=\bl{EXP}(im|x)
=\bl C(m|x)+i\ga_k\bl S^k(m|x)
\cr
\end{eqn}The spacelike behaviour is decisive for the probability interpretation -
here the indefinite metric has to be avoided.
In- and outgoing particles are a tangent spacelike phenomenon. 

The spacelike behaviour of particles is described 
explicitely by 
\begin{eqn}{l}
{\scriptsize\pmatrix{\bl C(m|x)\cr i\bl S^k(m|x)\cr}}
={m^2\over4\pi^2}\ep(m)m {\scriptsize\pmatrix{
-1\cr 4 {\p\over\p m x_k}\cr}}
{\p\over\p m^2x^2}
\brack{\log{|m^2x^2|\over4}\cl E_0(m^2x^2)-\cl F_0(m^2x^2)}\cr
\end{eqn}In addition to the 
functions $\cl E_n$, given in subsection 1.2, it involves the functions 
\begin{eqn}{l}
 \cl F_n(\xi^2)=n!{\SUM_{j=0}^\infty}{ (-{\xi^2\over4})^j\over j!(j+n)!}
\brack{\phi(j+n)-2\ga_0)},~j\ge1:~
\phi(j)=1+{1\over2}+\dots{1\over j}\cr
\ga_0={\phi(0)\over2}=\lim_{j\to\infty}\brack{\phi(j)-\log j}=0.5772...
\hbox{ (Euler's constant)}\cr
\end{eqn}

For relativistic ti\-me\-spa\-ce $\D(2)$,
the   quantization distributions $\bl c^k(m|x)$, $\bl s(m|x)$
and $\bl{exp}(im|x)$,  on the one side,  and 
the  Fock functions $\bl C(m|x)$, $\bl S^k(m|x)$  and $\bl{EXP}(im|x)$,
on the other side, are
completely different whereas their analogues coincide for time $\D(1)$ 
(subsection 1.1)
\begin{eqn}{rl}
{\scriptsize\pmatrix{
\cos m t\cr
i\sin mt\cr}}=
\int dq_0
\de(m^2-q_0^2)~\ep(q_0)~ 
{\scriptsize\pmatrix{q_0\cr m\cr}}
 e^{iq_0t}=
\int dq_0
\de(m^2-q_0^2)~\ep(m) {\scriptsize\pmatrix{m\cr q_0\cr}}
 e^{iq_0t}\cr
\end{eqn}The Fock functions of the position-momentum pair for
irreducible causal re\-pre\-sen\-ta\-tions in $\U(1)$
are exemplified by the harmonic oscillator with intrinsic length
$\ell^4={1\over\ka M}$
\begin{eqn}{rl}
\angle{ \acom{\ro u^\star}{\ro u}}(t)&= e^{imt}\cr
{\scriptsize\pmatrix{
\angle{\acom{iP}X}& \angle{\acom{X}X}\cr
\angle{\acom{P}P}&\angle{\acom{X}{-iP}}\cr}}(t)
&={\scriptsize\pmatrix{ 0&\ell^2\cr {1\over\ell^2}&0\cr}}
{\scriptsize\pmatrix{
\cos m t&i\ell^2\sin m t\cr 
{i\over\ell^2} \sin m t&\cos m t\cr}}\cr
\end{eqn}The value  for $t=0$ reflects the $\U(1)$-scalar product
for the crea\-tion-an\-ni\-hi\-la\-tion operator pair $(\ro u,\ro u^\star)$
\begin{eqn}{l}
\angle{\ro u^\star\ro u}=1,~\angle{\ro u\ro u^\star}=0\then
2\angle{ X^2}=\ell^2,~~~ 2\angle{P^2} ={1\over\ell^2}\cr
\end{eqn}

Going from time translations $\R(1)$ to 
relativistic  ti\-me\-spa\-ce translations $\R(2)$, the point $\R(1)_\Space=\{0\}$ 
(presence) is 'blown up' into the spacelike
submanifold $\R(2)_\Space$. 
For particle fields, 
i.e. with $\U(1)$-re\-pre\-sen\-ta\-tions
of the translations,  e.g. for the chiral components of a Dirac
electron field 
$\bl \Psi(x)={\scriptsize\pmatrix{\bl l(x)\cr\bl r(x)\cr}}$,
the creation and annihilation operators for the particle 
$(\ro u(\rvec q),\ro u^\star(\rvec q))$ and
antiparticle $(\ro a(\rvec q),\ro a^\star(\rvec q))$ involved 
\begin{eqn}{rl}
\bl l(x)&= \int{d^3q\over (2\pi)^3}
\sqrt{{|m|\over|q_0|}}
 ~e^{{\rvec q\over m}\rvec \si}~
{e^{ixq}\ro u(\rvec q)+ e^{-ixq}\ro a^\star(\rvec q)\over\sqrt2}\cr
\bl r(x)
&=  \int{d^3q\over (2\pi)^3}~\sqrt{{|m|\over|q_0|}}
e^{-{\rvec q\over m}\rvec \si}~
{ e^{ixq}\ro u(\rvec q)-e^{-ixq}\ro a^\star(\rvec
q)\over\sqrt2}\cr
\hbox{with}& q=(|q_0|,\rvec q),~~
q_0=\sqrt{m^2+\rvec q^2}\cr
\end{eqn}have the $\U(1)$-scalar product
\begin{eqn}{l}
\angle{\ro u^\star(\rvec p)\ro u(\rvec q)}=
\angle{\ro a^\star(\rvec p)\ro a(\rvec q)}=
(2\pi)^3\de(\rvec q-\rvec p)\cr
\angle{\ro u(\rvec p)\ro u^\star(\rvec q)}=
\angle{\ro a(\rvec p)\ro a^\star(\rvec q)}=0\cr
\end{eqn}

A rest system, used for the experiments, defines a  decomposition
of $\R(2)$ into  time and  space translations and a 
Lebesque measure  factorization $d^4q=d^3 qdq_0$.
A corresponding re\-pre\-sen\-ta\-tion of the 
spacelike behaviour shows the spherical waves for the particles
\begin{eqn}{l}
\bl C(m|\rvec x)={m^2\over(2\pi)^2}\int dq_0\vth(q_0^2-m^2)
{\sin \sqrt{q_0^2-m^2}|\rvec x|\over  |m\rvec x|}
,~~\int d^3 x\bl C(m|\rvec x)=1
\end{eqn}

Only the $\U(1)$-re\-pre\-sen\-ta\-tions of the  translations
have a positive definite scalar product as expressed by the Fock state.
The indefinite unitary re\-pre\-sen\-ta\-tions
have an indefinite inner product (subsection 3.2), as characterized by
${\scriptsize\pmatrix{1&0\cr 0&-1\cr}}$ for $\U(1,1)$.
Here the abelian $\U(1)$ Fock form
is inappropriate. 
It has to be replaced by the 
nonabelian Heisenberg-Killing  form  \cite{S912} leading to
trivial spacelike contributions in the Feynman propagator.

In the nondecomposable point measure approximation for the fundamental fields
for the realization of the ti\-me\-spa\-ce manifold only the $\U(1)$-re\-pre\-sen\-ta\-tion
of the translations can
give rise to in- and outgoing particles. This is effected by  
the familiar Feynman integration prescription for a particle
at its pole in the complex energy-momenta plane using
a limit vanishing imaginary part $m_1^2-io$ (subsection 1.2)
\begin{eqn}{l} 
 \angle{\cl C{ \ro b^\x}{ \ro b}}(x) 
={i\over \pi}
\int {d^4q\over\pi^2} {q \over(q^2-m_1^2+io)(q^2-m_2^2)_P^2}
e^{iqx}
\end{eqn}The re\-pre\-sen\-ta\-tions of 
the translations which cannot be spanned by
energy-momentum eigenvectors,
i.e. related to the reducible, but nondecomposable representations
- here the dipole - are confined to the causal submanifold
with the principal  value integration.  
 They  give rise to interactions, not,
however, to particles.


\begin{thebibliography}{99}

\bibitem{BRS}{C. Becchi, A. Rouet, R. Stora, {\it Ann. of Phys.} 98 (1976), 287}

 
\bibitem{BOE}{H. Boerner, {\it Darstellungen von Gruppen} (1955), Springer,
      Berlin, G\"ot\-tin\-gen, Heidelberg}

\bibitem{LIE13}{N. Bourbaki, {\it Lie Groups and Lie Algebras, Chapters 1-3} (1989),
Springer, Berlin, Heidelberg, New York, London, Paris, Tokyo}


\bibitem{BRAROB}{O. Bratelli, D.W. Robinson, 
{\it Operator Algebras and Quantum Statistical Mechanics 1} (1979)
Springer-Verlag, Berlin etc.}

 
\bibitem{FI}{D.R. Finkelstein, {\it Quantum Relativity} (1996),
Springer-Verlag, Berlin etc.}

\bibitem{GELNAI}{I.M. Gelfand, M.A. Neumark,
{\it Unit\"are Darstellungen der klassischen Gruppen}
(1950, German translation 1957), Akademie Verlag, Berlin}

\bibitem{GELVIL}{I.M. Gel'fand, M.I. Graev, N.Ya. Vilenkin,
{\it Generalized Functions V (Integral Geometry and Representation Theory)}
(1962, English translation 1966), Academic Press, New York and London}

 

\bibitem{HEI}{W. Hei\-sen\-berg, {\it Einf\"uhrung in die einheitliche
      Feldtheorie der Ele\-men\-tar\-teil\-chen} (1967), Hirzel, Stuttgart}

\bibitem{HEL}{S. Helgason, 
{\it Differential Geometry, Lie Groups and Symmetric Spaces}
(1978) Academic Press, New York etc.}
 
\bibitem{HIL}{J. Hilgert, G. Olafsson, {\it Causal Symmetric Spaces} (1997),
Academic Press, New York etc.}


\bibitem{NAK}{N. Nakanishi, I. Ojima, {\it Covariant Operator Formalism of
Gauge Theories and Quantum Gravity} (1990), World Scientific, Singapore etc.}
 
 \bibitem{NAIM}{M.A. Neumark,
{\it Lineare Darstellungen der Lorentzgruppe}
(1958, German translation 1963), VEB Deutscher Verlag der Wissenschaften,
Berlin}


\bibitem{RIC}{C.E. Rickart, {\it General Theory of Banach Algebras} (1960),
 D.van Nostrand, Princeton }

\bibitem{SHELO}{D.P. Shelobenko, {\it Doklady Akad. Nauk SSSR}
121 (1958), 586; 126 (1959), 482 and 935}


\bibitem{SNED}{I.N. Sneddon, 
{\it Special Functions of Mathematical Physics and Chemistry} (1961),
 Oliver and Boyd, Edinburgh and London} 

\bibitem{WEIZWP}{C.F.v. Weizs\"acker, {\it Zum Weltbild der Physik} (1993), 
Hirzel, Stuttgart}

 
 
\bibitem{WIG}{E. P. Wigner, {\it Annals of Mathematics} 40 (1939), 149}
 

\bibitem{S89}{H. Saller, {\it Nuovo Cimento} 104B (1989), 291}
 %%%%On the Nondecomposable Time Representations in Quantum Theories

\bibitem{S911}{H. Saller, {\it Nuovo Cimento} 104A (1991), 493}
%%% An Algebraic Interpretation of Quantized Gauge Interactions
%%% and BRS Transformations

\bibitem{S912}{H. Saller, {\it Nuovo Cimento} 106B (1992), 1319}
%%% Indefinite Metric and Asymptotic Probabilities

\bibitem{S921}{H. Saller, 
{\it Nuovo Cimento} 105A (1992), 1745 and 106A (1993), 1189}
%%% On the Isospin-Hypercharge Connection
%%% The Winding Numbers of the Standard Model,


 
\bibitem{S922}{H. Saller, {\it Nuovo Cimento} 108B (1993), 603
and 109B (1993), 255}
%%% Quantum Algebras I - Basic Algebraic and Topological Structures
%%% Quantum Algebras II - Tangent and Gauge Symmetries
 
 
 
\bibitem{S923}{H. Saller, 
{\it Nuovo Cimento} 106A (1993), 469}
 %% The Probability Structure and BRS Invariance of Noncompact Hamiltonians

 
\bibitem{SBH95}{H. Saller,  R. Breuninger and M. Haft
{\it Nuovo Cimento} 108A (1995), 1225}
%%% Quantum Fields \'a la Sylvester and Witt.

\bibitem{S96}{H. Saller,
{\it Intern. Journal of Theor. Phys.}, to be published}
%%% Analysis of Time-Space Translations in Quantum Fields


 \end{thebibliography}
\end{document}